# Efficient mid-term forecasting of hourly electricity load using generalized additive models

Monika Zimmermann[a,*], Florian Ziel[a]

[a] *Chair of Environmental Economics, esp. Economics of Renewable Energy*
*University of Duisburg-Essen*
*Germany*

**Abstract**

Accurate mid-term (weeks to one year) hourly electricity load forecasts are essential for strategic decision-making in power plant operation, ensuring supply security and grid stability, planning and building energy storage systems, and energy trading. While numerous models effectively predict short-term (hours to a few days) hourly load, mid-term forecasting solutions remain scarce. In mid-term load forecasting, capturing the multifaceted characteristics of load, including daily, weekly and annual seasonal patterns, as well as autoregressive effects, weather and holiday impacts, and socio-economic non-stationarities, presents significant modeling challenges. To address these challenges, we propose a novel forecasting method using Generalized Additive Models (GAMs) built from interpretable P-splines that is enhanced with autoregressive post-processing. This model incorporates smoothed temperatures, Error-Trend-Seasonal (ETS) modeled and persistently forecasted non-stationary socio-economic states, a nuanced representation of effects from vacation periods, fixed date and weekday holidays, and seasonal information as inputs. The proposed model is evaluated using load data from 24 European countries over more than 9 years (2015-2024). This analysis demonstrates that the model not only has significantly enhanced forecasting accuracy compared to state-of-the-art methods but also offers valuable insights into the influence of individual components on predicted load, given its full interpretability. Achieving performance akin to day-ahead Transmission System Operator (TSO) forecasts, with computation times of just a few seconds for several years of hourly data, underscores the potential of the model for practical application in the power system industry.

*Keywords:* Electricity Demand, Error-Trend-Seasonal Model, GAM, Holidays, Trend Behavior, Weather

***Nomenclature***

*List of Abbreviations*

**AIC** *Akaike Information Criterion*

**AR** *Autoregressive*

**ENTSOE** *European Network of Transmission System Operators for Electricity*

*Corresponding author
*Email addresses:* Monika.Zimmermann@uni-due.de (Monika Zimmermann), Florian.Ziel@uni-due.de (Florian Ziel)

**ETS** *Error-Trend-Seasonal*

**GAM** *Generalized Additive Model*

**GCV** *Generalized Cross Validation*

**GDP** *Cross Domestic Product*

**HldF** *Fixed date Holiday*

**HldP** *Holiday Period*

**HldW** *Weekday Holiday*

**HoD** *Hour of the Day*

**HoW** *Hour of the Week*

**HoY** *Hour of the Year*

**MAE** *Mean Absolute Error*

**REML** *Restricted Maximum Likelihood*

**RMSE** *Root Mean Square Error*

**Temp** *Temperature*

**TMAE** *Trace Mean Absolute Error*

**TSO** *Transmission System Operator*

*List of Symbols*

$\mathcal{N}(\mu, \sigma)$*: normal distribution with mean* $\mu$ *and standard deviation* $\sigma$

$\mathbb{N}$*: space of natural numbers*

$\mathbb{R}$*: space of real numbers*

$\mathcal{C}^2$*: space of two times continuously differentiable functions*

$\bigotimes$*: tensor product*

$\cup$*: union operator for sets*

## 1. Introduction

Accurate mid-term electricity load forecasts (weeks to one year), specifically, of an hourly resolution are crucial for all stakeholders in the power system industry. They guide electricity utilities in production and maintenance planning, ensuring grid stability and uninterrupted supply. They provide insights for the optimal sizing of energy storage systems. They are necessary for energy trading and investment firms to make informed decisions. They are instrumental in facilitating policymakers to a sensitive and socially acceptable transition from fossil fuels to electric power sources, particularly in sectors like transportation and heating.

For these tasks, the need for hourly resolution forecasts arises from the dynamics of the electricity market. Electricity markets commonly operate at hourly or even finer resolution accounting for the inter-day fluctuations in electricity demand and supply. On the demand side, these are attributed to our western society's hour-specific daily living and working behavior. On the supply side, they are caused by an increased share of renewables in the market. Consequently, coarser-resolution forecasts fail to meet the required accuracy for stakeholders to effectively fulfill their tasks in the power system industry, as argued by Behm et al. (2020); Agrawal et al. (2018).

While extensive research exists on hourly short-term (hours to a few days) load forecasting, the overview by Verwiebe et al. (2021) or Davis et al. (2016) suggests that literature remains scarce for hourly mid-term forecasts. This gap can be attributed to the central role of weather, in particular air temperature, in driving electricity demand contrasted by the lack of exact weather forecasts for a medium horizon, see e.g. Bashiri Behmiri et al. (2023); Kalhori et al. (2022); Agrawal et al. (2018); De Felice et al. (2015); Ghiassi et al. (2006), and the multifaceted nature of load time series. Following a classification proposed by Pierrot & Goude (2011), the various facets underlying load can be consolidated into three main groups:

(i) **Calender Characteristics**: These encompass the calender-based behavior of modern western societies entailing repetitive patterns of different seasonalities in electricity load, e.g. yearly patterns (higher load levels during winter vs. lower load levels during summer), weekly patterns (higher load levels on weekdays vs. lower load levels on weekends) and daily patterns (higher load during the day vs. lower load during the night), and holiday patterns.

(ii) **Weather Characteristics**: These include air temperature, humidity, cloud cover and wind speed that affect load due to their impact on electric heating and lighting.

(iii) **Socio-Economic Characteristics**: These involve macroeconomic, socioeconomic and energy variables, e.g. economic growth, industrial production, population size or fossil fuel prices, which influence, in particular, in their unit root behavior medium to long-term levels of load.

Proposed methods in mid-term load forecasting range from classical time series, see e.g. Petropoulos et al. (2022) for a concise overview, to sophisticated machine learning models based on e.g. different types of neural networks, gradient boosting machines or random forests, see e.g. the highly cited winning method of a EUNITE Competition Chen et al. (2004) or Bashiri Behmiri et al. (2023); Zhang et al. (2023); Li et al. (2023); Kalhori et al. (2022); Butt et al. (2022); Oreshkin et al. (2021); Han et al. (2019); Agrawal et al. (2018); Bouktif et al. (2018). Furthermore, hybrid models combining diverse forecasting methods have emerged lately, as summarized by Petropoulos et al. (2022) and Hong et al. (2016). For instance, winning methods in the IEEE DataPort Competition, see De Vilmarest & Goude (2022); Ziel (2022), employed combinations of different techniques, including Generalized Additive Models (GAMs). Notably, in De Vilmarest & Goude (2022), GAMs are among the models with the lowest forecasting errors compared to e.g. autoregressive, linear regression and Multi-Layer Perceptron models.

Similarly, GAMs have consistently demonstrated outstanding performance in previous forecasting competitions, such as achieving top-three results, see Nedellec et al. (2014), in GEFCom 2012 Davis et al. (2014), and securing the top two spots, see Gaillard et al. (2016); Dordonnat et al. (2016), in GEFCom 2014 Hong et al. (2016). Motivated by the success of GAMs in load forecasting competitions alongside their effectiveness in early load forecasting studies, see e.g. Pierrot & Goude (2011); Fan & Hyndman (2012); Goude et al. (2014), and Europe-wide studies, see Gurriaran et al. (2023); Troccoli et al. (2018), we adopt this framework as the foundation for our proposed load forecasting model.

GAMs offer a distinctive blend of interpretability and efficient estimation while capturing non-linear relationships. This stems from their underlying linearity, i.e. a linear model structure, in non-linear functions. This combination of attributes makes them a compelling choice when computational efficiency and model understanding are crucial and consequently suitable as weak learners, e.g. as input for neural networks and in ensembling methods, see Lepore et al. (2022).

Specifically, we propose a two-stage GAM built from interpretable P-splines that is enhanced with autoregressive post-processing. This model uses exponentially smoothed temperatures, Error-Trend-Seasonal (ETS) modeled non-stationary states, a nuanced representation of holiday effects with weekday variations and seasonal information as inputs. With this approach, we contribute five key innovations to the field of mid-term hourly load forecasting:

(i) **Accurate, Fast and Interpretable Model:** We develop a GAM-based forecasting model utilizing shrinkage and discretization that covers all relevant characteristics of load while maintaining computation times of a few seconds for hourly year-ahead forecasts. The model is interpretable in each input characteristic, allowing us to derive the effect of holidays and weather inputs. It has superior forecasting accuracy compared to state-of-the-art methods.

(ii) **Individual Holiday Modeling:** We introduce a novel modeling approach that captures the effect induced by each holiday individually through a holiday-specific activation variable. This approach enables the modeling of holidays limited to a specific region of a country as well as holiday periods such as the Christmas/New Year vacation period. Moreover, we distinguish the effects of holidays that occur on a fixed date (e.g. New Year's Day) from those that occur on a fixed weekday (e.g. Good Friday).

(iii) **Mid-Term Temperature Effects:** We show that the incorporation of temperature information improves the mid-term forecasting accuracy significantly even though our simple mid-term temperature forecasts provide essentially only seasonal patterns.

(iv) **Mid-Term State Components:** To account for the mid-term gradual change of load levels due to economic or socioeconomic factors, we incorporate ETS modeled state components in our model that have a persistent trend behavior.

(v) **Comprehensive Robustness Check:** The robustness of the model is demonstrated through an extensive evaluation across 24 European countries[1], illustrated by the map in Figure 1, over more than 9 years (2015-2024), including volatile periods due to the COVID-19 pandemic and the energy crisis caused by the Russian invasion of Ukraine.

[1] While data from all European countries participating in the European Network of Transmission System Operators for Electricity (ENTSOE) was initially collected, some countries were ultimately excluded from the analysis due to data quality issues.

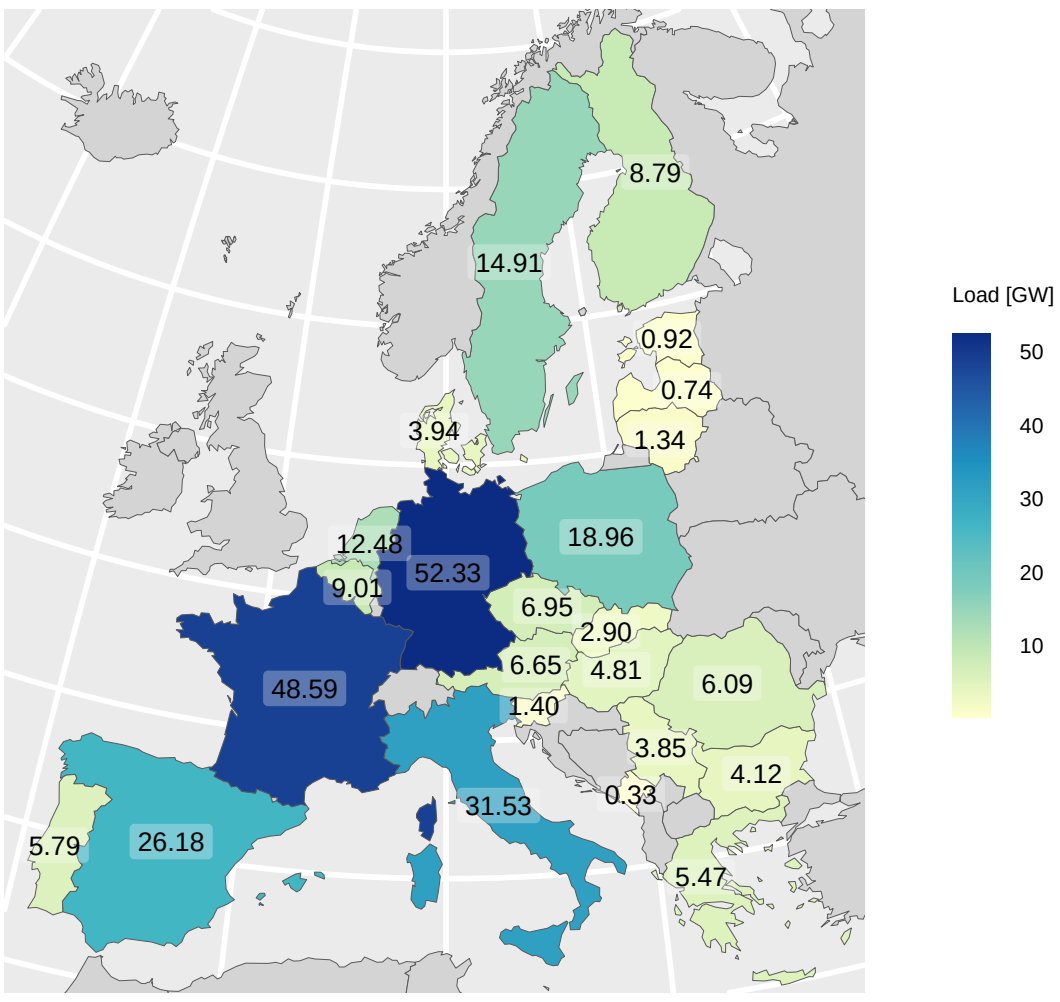

Figure 1: Average load of 2023 in the 24 European countries considered in the study.

In a forecasting study, we found that our proposed model has superior accuracy compared to a set of 10 common benchmark models and compared to the same model excluding temperature, holiday and autoregressive effects. Regarding the inclusion of the ETS state components improvement is diminished in many countries during the first lockdowns due to COVID. Incorporating exogenous information on COVID-19 lockdowns, thus relaxing the assumption of a persistent trend, enhanced forecasting accuracy during the pandemic.

For a quarter of the considered countries, accuracy reaches the range of the short-term day-ahead Transmission System Operator (TSO) forecasts which highlights the applicability of our model in the electricity utilities industry.

The remainder of this paper is organized as follows: Section 2 details the multifaceted characteristics of load time series. Section 3 introduces the applied statistical models, with a concise background on GAMs and additive autoregressive models in 3.1 and 3.2, respectively. Section 4 describes our methodology beginning with a conceptual overview of our model in 4.1, data preprocessing in 4.2 and specification, estimation and forecasting of our model in 4.3-4.4. Benchmark models are detailed in Subsection 4.5 and the forecasting study and evaluation design in Section 5. The results of the comprehensive forecasting study are discussed and interpreted in Section 6. Finally, Section 7 concludes and outlines further research to be built upon this work.

## 2. Data

### *2.1. Electricity Load*

The load data used for this study were retrieved from ENTSOE, cover the period from January 1st, 2015 to February 17th, 2024 and are available for all considered countries in an hourly resolution. This data incorporates implicit calendar and time information, including daily, weekly and annual seasonalities, as well as changes due to daylight saving time and leap years.

The load profiles exhibit strong seasonal patterns across daily, weekly and annual timescales, as shown by Figures 2 and 3. These patterns include lower loads during summer compared to winter months, higher loads on weekdays compared to weekends and lower loads at night compared to daytime. Consequently, autocorrelation is high for the most recent lags, lags of the neighboring hours on the previous day and the previous week. Notably, yearly seasonal variations are more pronounced in France due to the country's reliance on electric heating.

Load is influenced by past observations not only in terms of its autoregressive components but also in terms of its previous mid-term level. More precisely, random shocks, e.g. due to a financial crisis, carbon tax, technological advancements, supply shortages or large-scale migration caused by war, on load can cause permanent deviations from a predetermined equilibrium level. Since electricity demand is strongly tied to the economy, load inherits the prevalence of these shocks via transmission flows from key macroeconomic, socioeconomic and energy variables Schneider & Strielkowski (2023); Hendry & Juselius (2000); Smyth (2013). Key macroeconomic drivers of electricity demand are, for instance, Gross Domestic Product (GDP), employment rate, Consumer Price Index and Industrial Production Index measuring economic well-being, purchasing power and production output of manufacturing. These macroeconomic variables, with GDP being a prime example, see Kalhori et al. (2022); Harvey et al. (2007), are generally recognized to exhibit persistent shifts in their level component. This has been, firstly, evidenced by the seminal work of Nelson & Plosser (1982) and mostly supported by revisiting studies, see e.g. Gil-Alaña & Robinson (1997); Perron (1997) among others. Socioeconomic and energy variables affecting load are, for instance, population size, see Troccoli et al. (2018), and fossil fuel prices or carbon taxes through industrial production. Similar to macroeconomic variables, these factors are predominantly recognized to exhibit a permanent shift in levels, see e.g. Schneider & Strielkowski (2023); Narayan & Liu (2015) and Smyth (2013) for a literature overview.

Prevailing shifts in the level component, defined in time series decomposition as the average value over a specific timeframe, see Svetunkov (2023) for details, contribute to the non-stationarity of load data. Recall that non-stationary time series exhibit fluctuations in their statistical properties, e.g. mean, variance and autocorrelation, over time. By differentiating the load time series, its level component can be stabilized. This form of non-stationarity is known as a unit root.

Following the illustration of a long-term level in Ziel (2019), Figure 2 depicts the hourly load, the weekly and yearly moving average load in France and Germany. In both countries, the averaged load data exhibit gradual shifts in particular in the last four years, emphasizing their non-stationarity.

Our study deliberately bypasses formal unit root testing due to the ongoing research in this area, see e.g. Schneider & Strielkowski (2023), and the statistical complexities inherent in applying such tests to electricity time series. These complexities stem from the presence of structural breaks, non-linearities and seasonalities at various frequencies in the data, and potential small sample biases when aggregating data to long-term, e.g. annual horizons. For a comprehensive assessment of unit root behavior in electricity time series, we recommend the literature reviews by Schneider & Strielkowski (2023); Smyth (2013).

### *2.2. Holiday Information*

Holidays, of specific days or extended periods, significantly impact human behavior, often due to legal regulations. This usually leads to decreased work and industrial activity, consequently reducing electricity demand[2]. This disruption to the typical weekly load pattern is evident in Figure 3, which shows the load time series for the Easter and Christmas holiday time in Germany. Notably, the load profile of a holiday on a weekday, e.g. May 1st in 2015, frequently resembles that of a weekend.

[2]Note that areas with high tourism may experience increased demand during holidays.

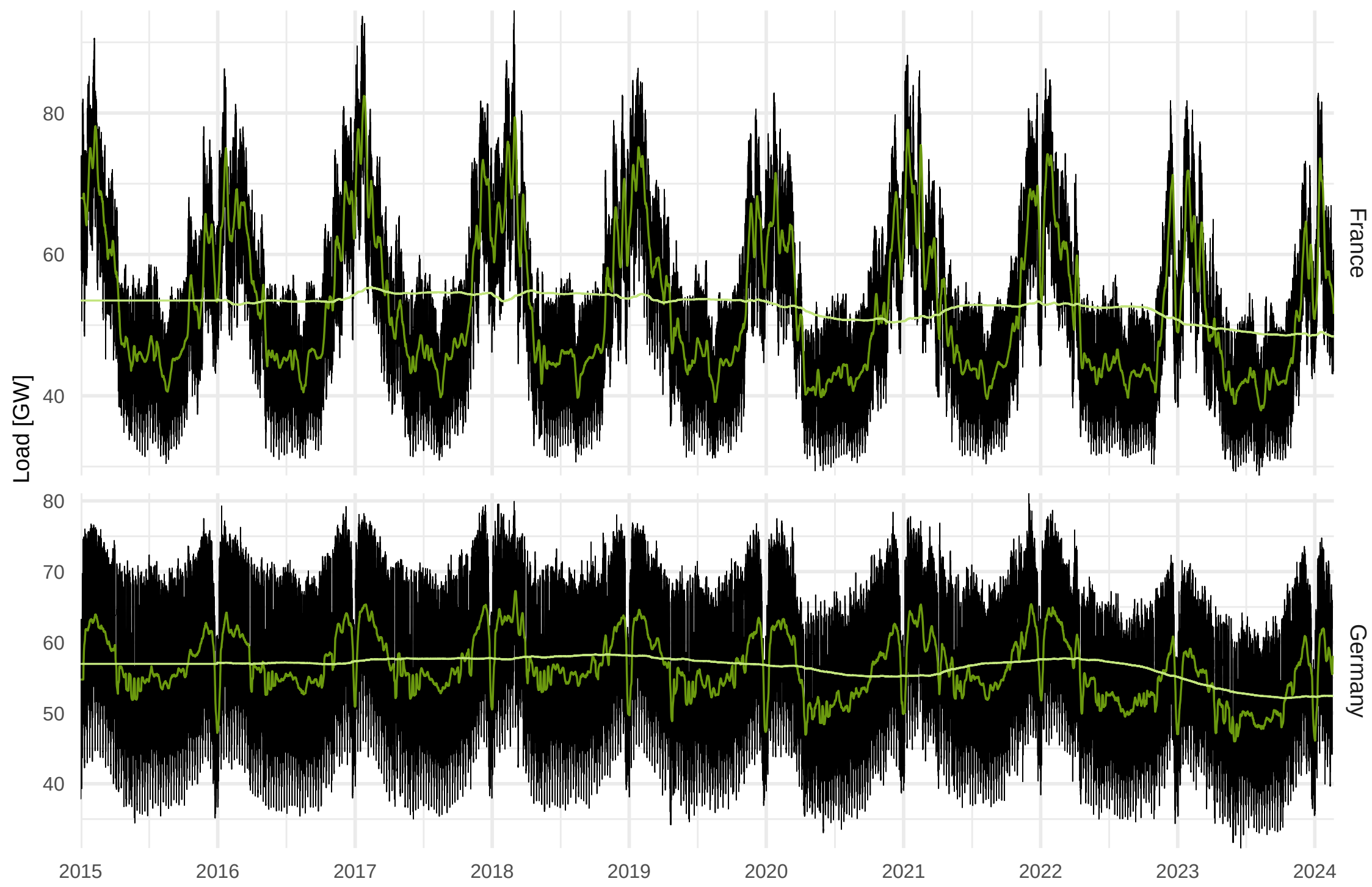

Figure 2: Hourly load (black), weekly (dark green) and yearly, i.e. 52 weeks, (light green) moving average load from January 1st, 2015 to February 17th, 2024.

Holiday information was collected from Nager.Date, see Hager (2024), is available from January 1st to December 31st of 2000 to 2030 and encompasses both the label and the day of occurrence of the holiday. These details undergo preprocessing, as described in Subsection 4.2, to obtain holiday variables as a function of time. While holiday information is known in advance, making them deterministic information based on the calendar, their impact on load can be challenging to capture. This is due to their infrequent occurrence, usually once a year, making them a rare event in the context of load forecasting Ziel (2018); Hong (2010).

Two common holiday categories exist, see Ziel (2018); Ziel & Liu (2016); Hong (2010):

- Weekday holidays: These holidays occur annually on a specific weekday (e.g. Good Friday), but their date varies within the year. They usually exhibit a stable impact on load each year.
- Fixed-date holidays: These holidays occur on the same date every year (e.g. New Year's Day), but the weekday on which they fall varies. They usually have a varying impact on load depending on the weekday on which they fall in a year.

In addition to these single-day categories, we consider a winter holiday period encompassing December 18th to January 6th in every country.

Figure 3 on load in the Easter and Christmas holiday time in Germany highlights the two distinct categories of single-day holidays. The Easter holidays Good Friday and Easter Monday have varying dates between late March and the middle of April. Their impact on load has remained consistent over the years. In contrast, the fixed date holiday Labour Day on May 1st exhibits varying effects on load depending on the weekday it falls on. Notably, Labour Day on Sunday in 2016 resulted in negligible load reduction. The Christmas period is particularly evident, with a reduced load between Christmas and New Year's Eve. For a similar Figure on load in France, we refer to Figure 15 in Appendix B. In France, the strong reliance of load on electric heating diminishes the visibility of holiday effects during the Christmas period.

### *2.3. Temperature*

Hourly temperature observations are collected for each considered country from meteostat.net (Meteostat Developers (2024)) starting from January 1st, 2015, to February 17th, 2024, at weather stations located in the most populous cities within each country. As load data, these temperatures exhibit clear seasonal patterns, including weekly and annual cycles, see Figure 4 on load and temperatures in France and Germany.

The non-linear relationship between temperature and load has been confirmed through various studies, see e.g. the overview by Verwiebe et al. (2021); Davis et al. (2016) and the comprehensive European study by Bessec & Fouquau (2008) or Bashiri Behmiri et al. (2023); Ziel (2018); Xie et al. (2018) among others. During summer when temperatures increase, load typically rises due to cooling demands. Conversely, during winter when temperatures fall, load increases in countries where electric heating is prevalent. As our model aims to encompass data from a diverse range of European nations, including countries with pronounced electric heating and cooling, like France, Spain and Italy, among others, factoring in temperature becomes essential for accurate load forecasting. For illustration, Figures 4 and 5 show the high sensitivity and non-linear relationship of load to temperature in France compared to a low sensitivity in Germany.

### *2.4. Countries*

The data described above were initially gathered for all European countries participating in ENTSOE. However, certain countries had to be excluded due to data quality concerns, such as more than 5% missing load data, unavailability of day-ahead load or weather data, or more than 5 blocks of consecutive zeros in reported load. This resulted in a final dataset encompassing data from 24 European countries: Austria, Belgium, Bulgaria, Czech Republic, Germany, Denmark, Estonia, Spain, Finland, France, Greece, Hungary, Italy, Lithuania, Latvia, Montenegro, Netherlands, Poland, Portugal, Romania, Serbia, Sweden, Slovenia, Slovakia. The average load of these countries in 2023 is illustrated by the map in Figure 1. The figures in this and the subsequent sections focus on France and Germany, as they are among the countries with the largest electricity demand and exhibit notable differences, e.g. in their dependence of load on temperature.

## 3. Primer On Statistical Models

### *3.1. Generalized Additive Models*

Generalized Additive Models (GAM) are the main building blocks to capture the multifaceted characteristics of load in our proposed forecasting model. In our application, a GAM describes a response

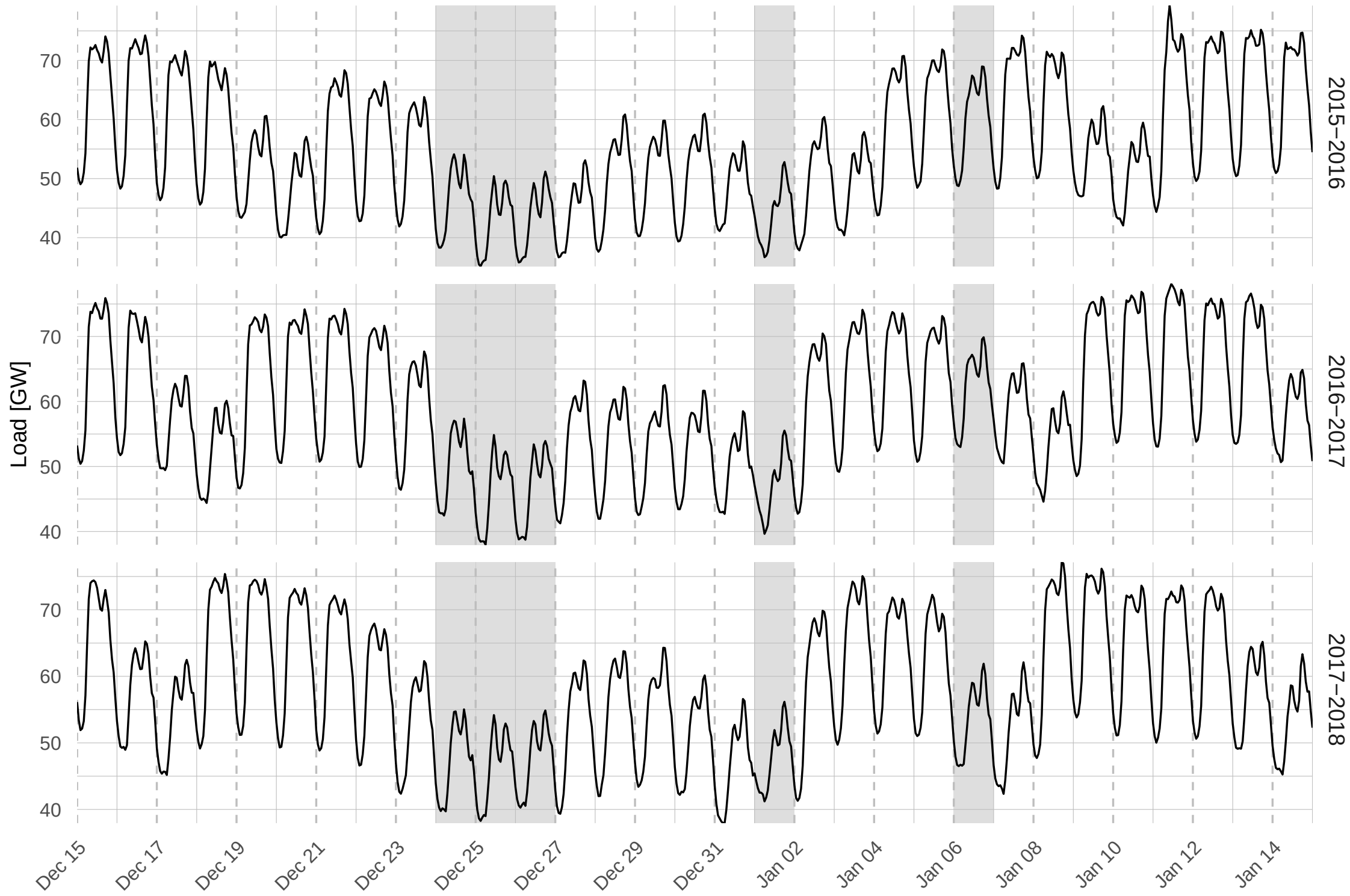

(a) December 15th in 2015, 2016 and 2017 to January 15th in 2016, 2017 and 2018.

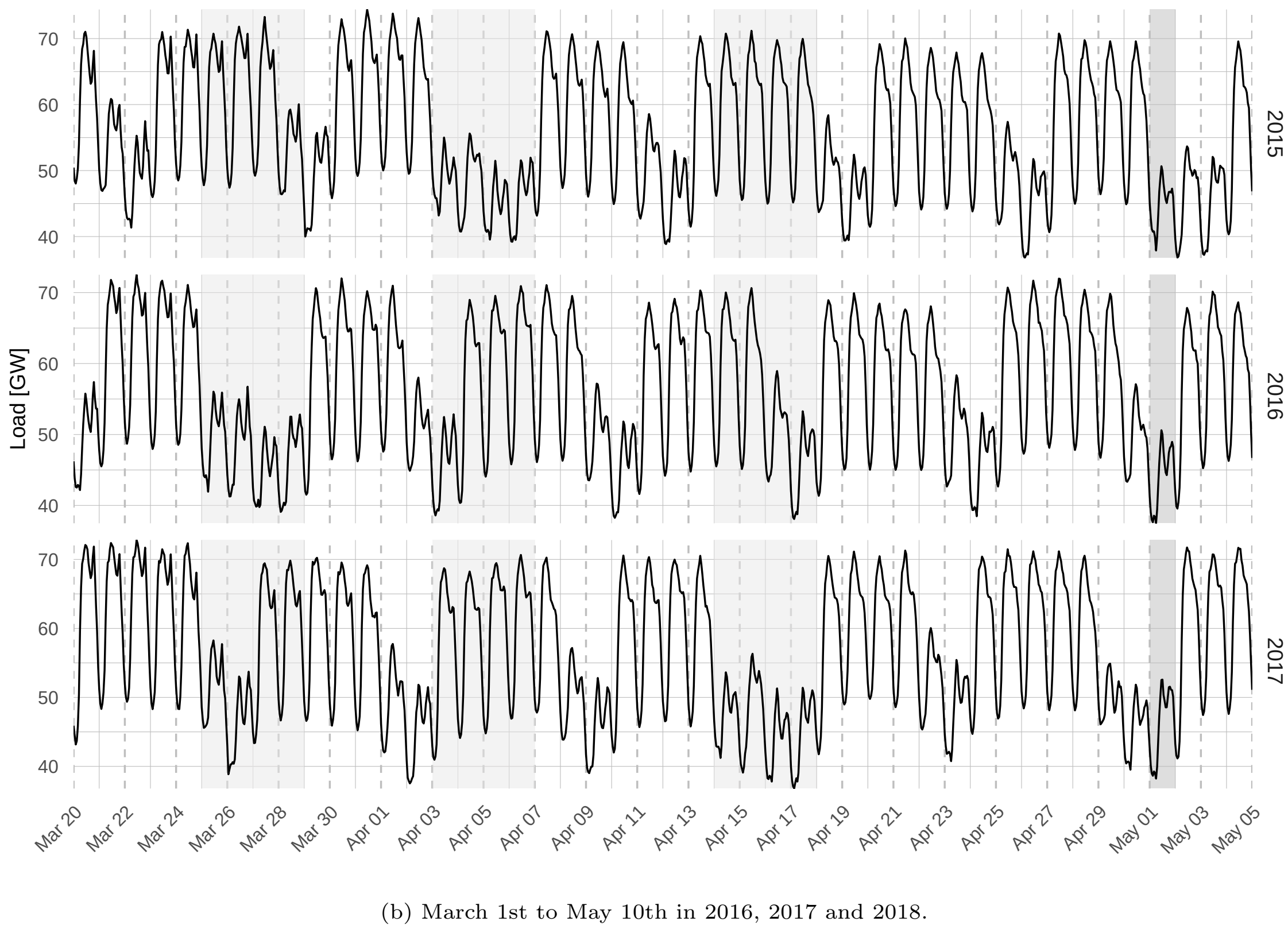

(b) March 1st to May 10th in 2016, 2017 and 2018.

Figure 3: Hourly load in Germany in the Easter and Christmas holiday time with weekday holidays shaded in light grey and fixed date holidays shaded in dark grey.

variable $Y_t$ as an additive combination $\mu$ of smooth functions in covariates $\boldsymbol{X}_t = (X_t^1, ..., X_t^m) \in \mathbb{R}^m$, $t \in \{1, .., T\}$. More precisely, there are smooth functions $f_i : \mathbb{R}^m \longrightarrow \mathbb{R}$ in the covariates $\boldsymbol{X}_t$ such that

$$Y_t = \mu(\boldsymbol{X}_t) + \varepsilon_t, \quad \mu(\boldsymbol{X}_t) = f_1(\boldsymbol{X}_t) + f_2(\boldsymbol{X}_t) + ... + f_M(\boldsymbol{X}_t) \tag{1}$$

for $\varepsilon_t \sim \mathcal{N}(0, \sigma)$. The objective to estimate in (1) is $\mu$ or the smooth functions[3] $f_1, ..., f_M$, respectively. Though in general, these smooth functions are arbitrary, we consider their marginals to lie in the $k_1, ..., k_m$-rank spans of $d$-degree polynomial spline basis functions $\{b_1^1, ..., b_{k_1}^1\}, ..., \{b_1^m, ..., b_{k_m}^m\}$:

$$f(x^1, ..., x^m) = \sum_{i=1}^{k_1} ... \sum_{j=1}^{k_m} \beta_{i,...,j} b_i^1(x^1) \cdot ... \cdot b_j^m(x^m). \tag{2}$$

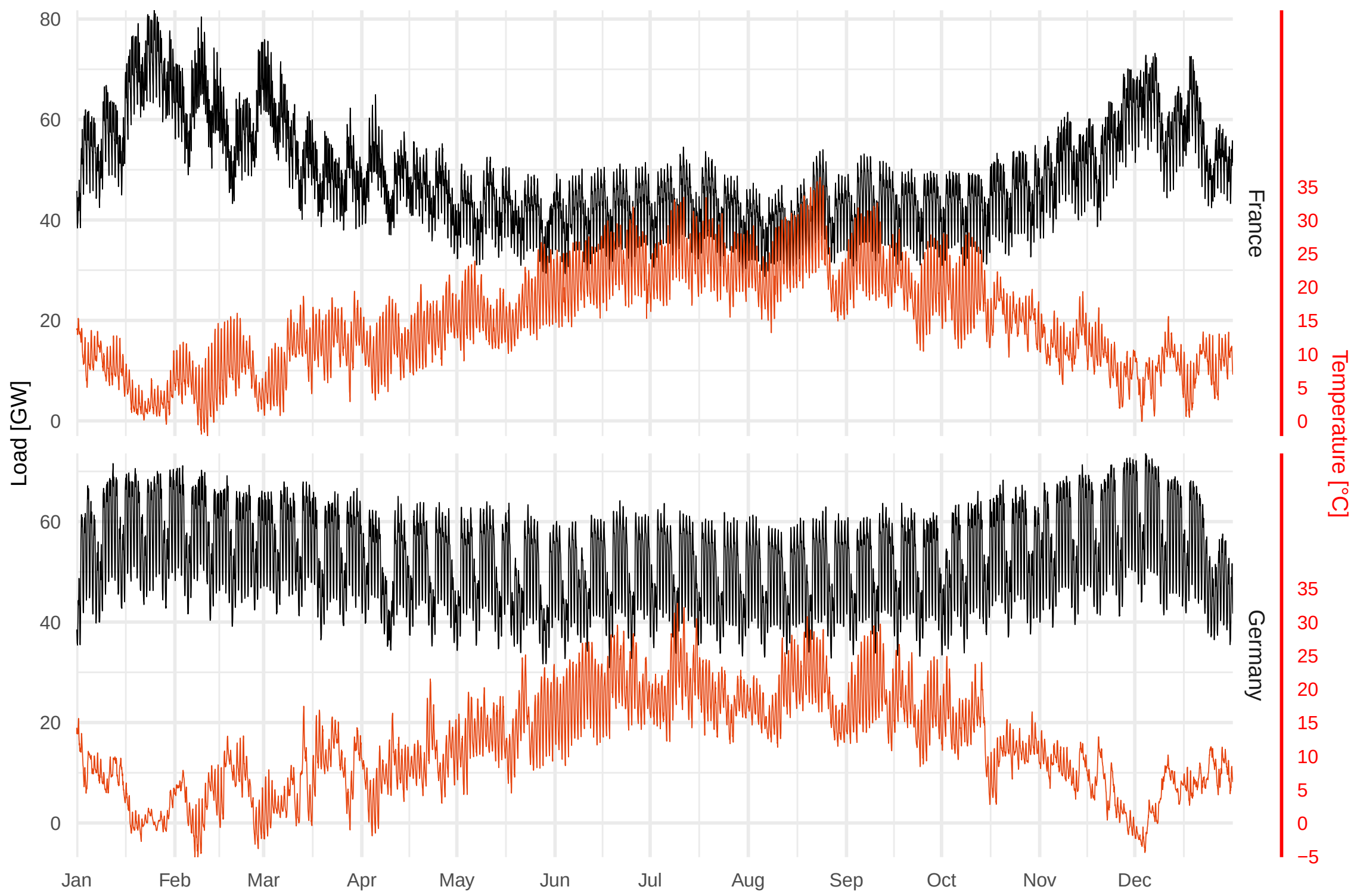

Figure 4: Hourly load (black) and temperature (red) in 2023.

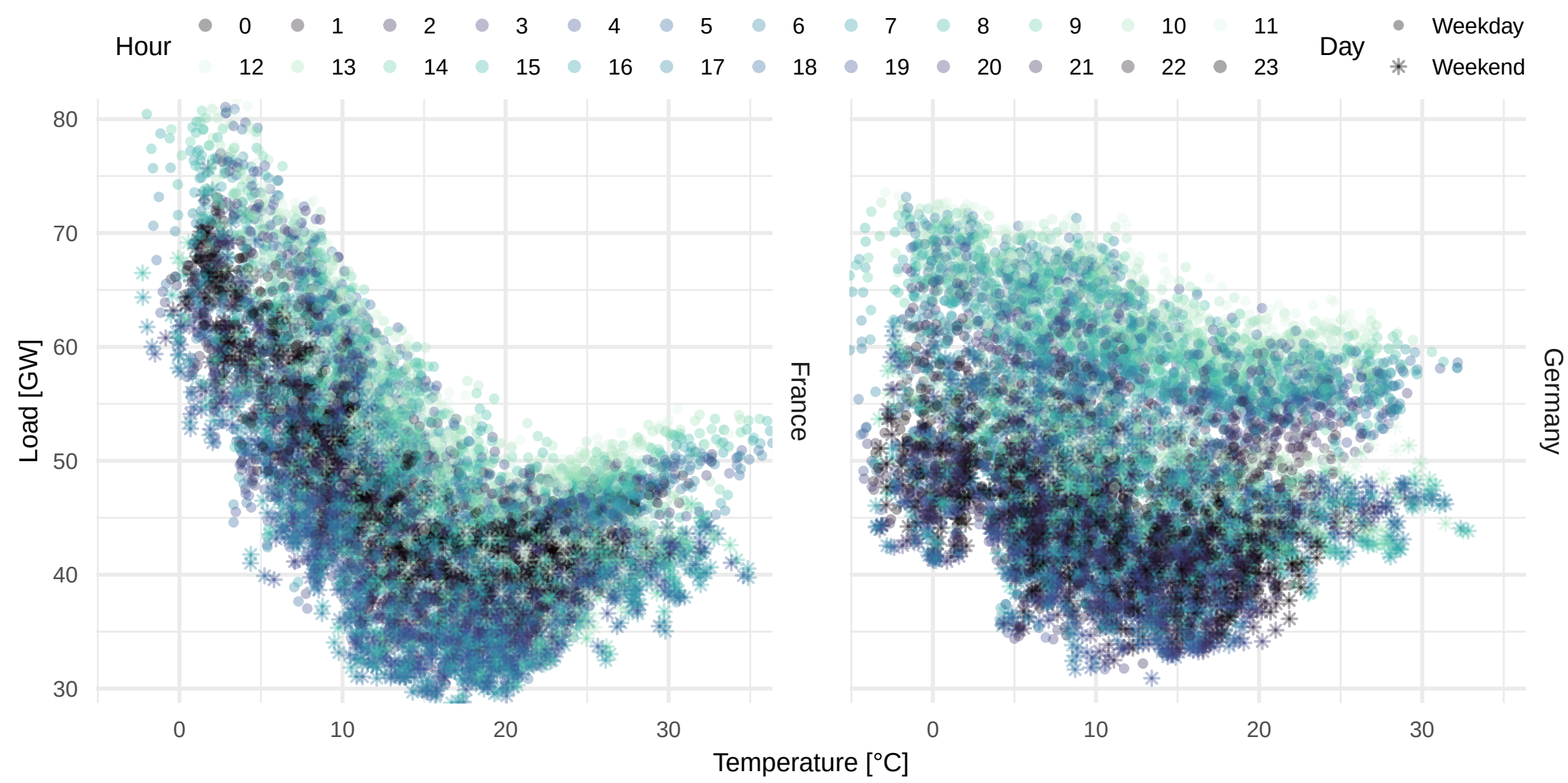

Figure 5: Temperature vs. hourly load in 2023.

The resulting curve $f : \mathbb{R}^m \longrightarrow \mathbb{R}$ becomes a spline of degree $d$ if the basis functions are joined together at a $m$-dimensional grid of knots $\{x_1^1, ..., x_{n_1}^1\} \times ... \times \{x_1^m, ..., x_{n_m}^m\}$ such that $f$ has finite support $[x_1^1, x_{n_1}^1] \times ... \times [x_1^m, x_{n_m}^m]$ and $f \in \mathcal{C}^2$. Representation (2) of a smooth function $f$ as a linear combination of the products of basis functions, allows the application of linear estimation methods, such as an ordinary least squares estimation on the evaluation of basis functions. The model matrix $\mathbb{X}$ in this case has rows of the form $\mathbb{X}_t = (\mathbb{X}_{t,1}^1 \otimes ... \otimes \mathbb{X}_{t,m}^1, ..., \mathbb{X}_{t,1}^M \otimes ... \otimes \mathbb{X}_{t,m}^M)$ for the tensor product $\otimes$ on matrix space and $\mathbb{X}_{t,j}^l = (b_1^j(X_t^j), ..., b_{k_j}^j(X_t^j))$ evaluating the marginal bases of the smooth term $f_l$ in covariate $X^j$ at time $t \in \{1, .., N\}$. Among various polynomial spline bases, the B-spline basis is particularly common, see Eilers & Marx (2021).

To smooth the terms $f_1, ..., f_M$ in (1) a penalized least-squares objective with smoothing parameters $\lambda_i \in \mathbb{R}_{\geq 0}^m$ is applied:

$$||\boldsymbol{Y} - \mathbb{X}\boldsymbol{\beta}||^2 + \sum_{i=1}^{M} \lambda_i \mathcal{P}(f_i). \tag{3}$$

Consequently, the smoothing parameters $\lambda_i$ balance an optimal model fit, achieved by low values of $\lambda_i$, and high levels of smoothness, achieved by high values of $\lambda_i$. In the case of B-splines, Eilers & Marx (2021) introduced a difference-based penalty. Specifically, they penalize the deviation in parameters $\beta_i$ for equidistant knot spacing:

$$\mathcal{P}\left(\sum_{i=1}^{k} \beta_i b_i(x)\right) = \sum_{i=1}^{k-p} (\Delta^p \beta_i)^2 = \boldsymbol{\beta}^T S \boldsymbol{\beta}, \tag{4}$$

for a one-dimensional term, where $\Delta$ is the difference operator. The resulting smooth terms are named *P-splines*. For a so-called cyclic P-spline, terms are added to penalize deviation between the first and last coefficients, e.g. $(\beta_1 - 2\beta_n + \beta_{n-1})^2$ and $(\beta_2 - 2\beta_1 + \beta_n)^2$ for $p = 2$. The generalization of $\mathcal{P}$ to multivariate splines penalizes parameter deviation individually for each marginal smooth, where remaining covariates are integrated out and integrals are approximated or derived as sums.

To estimate the smoothing parameters $\lambda_i$, criteria such as the unbiased risk estimator, the generalized cross validation (GCV) score and variations thereof can be minimized. Alternatively, Restricted Maximum Likelihood (REML) estimation in a mixed model view of smoothing can be applied. For these approaches differently efficient iteration schemes with estimation of $\beta_i$ by minimization of (3) can be applied. For further insights on GAM estimation, we refer to Wood (2017) or Lepore et al. (2022). GAM modeling and estimation methods are implemented in the R-package mgcv, see Wood (2017).

[3]Note that smooth functions $f_1, ..., f_M$ in (1) can only be identified and thus estimated up to a constant. Consequently, practical applications employ identifiability constraints, commonly $\sum_{t=1}^{T} f_i(\boldsymbol{X}_t) = 0$ or $f_i(0) = 0$, see Wood (2017) for details.

### *3.2. Additive Autoregressive and State Models*

To capture the effect on load by past observations two types of models, which generate forecasts as weighted sums of autoregressive and initial state components, are incorporated in our load forecasting model: Autoregressive (AR) and Error-Trend-Seasonal (ETS) Models.

For AR models (see Tab. 1, col. 1) the optimal autoregressive order $p$ is chosen by minimizing Akaike information criterion (AIC) over $p = 1, \dots, p_{\max}$ and parameter estimation is carried out by the Burg method, see Brockwell & Davis (2016).

While for AR models no assumption is imposed on the weights assigned to past observations, additive ETS models (see Tab. 1, col. 2-3), formulated by Hyndman & Athanasopoulos (2018), impose exponentially diminishing weights. This is obtained by applying exponential smoothing with smoothing equations for the additive decomposition of a time-series $Y_t$ into unobserved states, i.e. level $l_t$ and seasonal $s_t$ component[4]. From the model formulation, it directly follows that the states and thus $Y_t$ itself exhibit a non-stationary behavior of unit root type, i.e. by first-order and seasonal differentiating a stationary white noise process remains. The evolution of this unit root is regulated by the smoothing parameters $\alpha, \gamma \in (0, 1)$.

For estimation, Svetunkov (2023) implemented different multistep loss-function approaches. Specifically, estimation by trace mean absolute error TMAE minimizes the sum of the $1, \dots, h_{\text{ETS}}$ steps ahead in-sample MAEs:

$$\text{TMAE} = \sum_{j=1}^{h_{\text{ETS}}} \frac{1}{T - h_{\text{ETS}}} \sum_{t=1}^{T-h_{\text{ETS}}} |e_{t+j|t}| \tag{5}$$

Table 1: AR model of AIC-selected order $p \in \{1, \dots, p_{\max}\}$, additive ETS models with level state $l_t$, seasonal state $s_t$, smoothing parameters $\alpha, \gamma \in (0, 1)$ and periodicity $m$ for zero mean white noise $\epsilon_t$ with forecasting equations for $h = 1, 2, \dots$.

| $\text{AR}_{\text{AIC}}(\phi; p_{\max})$ | $\text{ETS}(\alpha)$ | $\text{ETS}(\alpha, \gamma; m)$ |
|---|---|---|
| $Y_t = \sum_{k=1}^{p} \phi_k Y_{t-k} + \epsilon_t$ | $Y_t = l_{t-1} + \epsilon_t$ | $Y_t = l_{t-1} + s_{t-m} + \epsilon_t$ |
| $\widehat{Y}_{t+h\|t} = \sum_{k=1}^{p} \phi_k \widehat{Y}_{t+h-k\|t}$ | $l_t = l_{t-1} + \alpha\epsilon_t$ | $l_t = l_{t-1} + \alpha\epsilon_t$ |
| $\widehat{Y}_{t+h-k\|t} = Y_{t+h-k}$ for $h \leq k$ | $\widehat{Y}_{t+h\|t} = l_t$ | $s_t = s_{t-m} + \gamma\epsilon_t$ |
| | | $\widehat{Y}_{t+h\|t} = l_t + s_{t+h-m\lceil \frac{h}{m} \rceil}$ |

## 4. Methodology

### *4.1. Modelling Concept*

As detailed in Subsection 2.1, accurately forecasting load is a challenging task due to its multifaceted characteristics. Load depends on deterministic information, i.e. the seasonalities of different periods and holidays, and is sensitive to temperature fluctuations. Moreover, it is influenced by past observations. Thus, autoregressive components and non-stationary components need to be incorporated into a forecasting model. To address these diverse characteristics, we propose a two-stage GAM model integrating

[4]In all ETS models considered in this work, the trend component is assumed to be zero.

ETS and AR models. The general process of the model is illustrated by the flowchart in Figure 6 and described in the subsequent.

Let $Y_t \in \mathbb{R}$ and $X_t^{\text{Temp}} \in \mathbb{R}$ be the electricity load and temperature at time $t$ for $t \in \{1, ..., T\}$, respectively. Additionally, we have access to deterministic information about seasonalities and holidays for the extended horizon $t \in \{1, ..., T+H\}$. It is the objective to forecast load over a mid-term prediction horizon $H$ of several weeks to one year, resulting in forecasts $\widehat{Y}_{T+1}, ..., \widehat{Y}_{T+H}$.

**(i) Temperature forecast**: In a first step, see Figure 6 on the left-hand side, the proposed model constructs temperature forecasts $\widehat{X}_t^{\text{Temp}}$, $t \in \{T+1, ..., T+H\}$ for the mid-term horizon $H$. Since accurate weather and satellite imagery-based temperature forecasts are not feasible for mid-term horizons, we rely on deterministic seasonalities, i.e. the time of the day and the time of the year, as modeling inputs. Observed temperatures $X_t^{\text{Temp}}$, $t \in \{1, ..., T\}$ are, first, preprocessed to a smoothed time series using an ETS model (see Tab. 1, col. 2) with fixed smoothing parameters. This removes fluctuations not capturable by deterministic seasonalities and creates more suitable inputs for the subsequent models to forecast temperature and load. A GAM model is, secondly, applied to the smoothed temperature. Recognizing the limitations of lagged components in GAMs for mid-term horizons, we employ an AR model on the residuals of this GAM. The forecasted temperature is obtained by summing the individual forecasts from both models. By this, we incorporate crucial short-term temporal dependencies while avoiding computationally expensive solutions, like multiple GAM estimations with differently recent lags.

**(ii) States fit and forecast**: In a second step, see Figure 6 middle part, the proposed model constructs level and seasonal states fit and forecast. For this purpose, firstly, load is modeled by a GAM from the smoothed temperatures obtained in (i) and deterministic seasonal and holiday information. Secondly, an ETS model (see Tab. 1, col. 3) with an annual seasonality is applied to this GAM's residuals and fitted states are retrieved. The forecasted states follow directly from these fits. To estimate the ETS model, we employ the TMAE calculated as the sum of in-sample MAEs over a multiple of the weekly period. By this, fluctuations occurring in a smaller period are smoothed and not mirrored by the fitted and forecasted states component, which is well-suited, particularly, in the context of mid-term forecasting.

**(iii) Load forecast**: Lastly, see Figure 6 right-hand side, load is forecasted. For this purpose, load is modeled again by a GAM with the smoothed temperatures, seasonal and holiday information and, additionally, with the ETS unit root states as inputs. This approach captures the unit root behavior in the load time series and mitigates the risk of the GAM inputs in (ii) adapting unduly to this non-stationary pattern. For the same rationale as in (i), autoregressive dependencies of the load time series are captured by a secondary AR model on the residuals. Finally, load is forecasted as the sum of the forecasts of these two models with the temperature, level and seasonal forecasts obtained in (i) and (ii) and the deterministic seasonal and holiday information over the horizon.

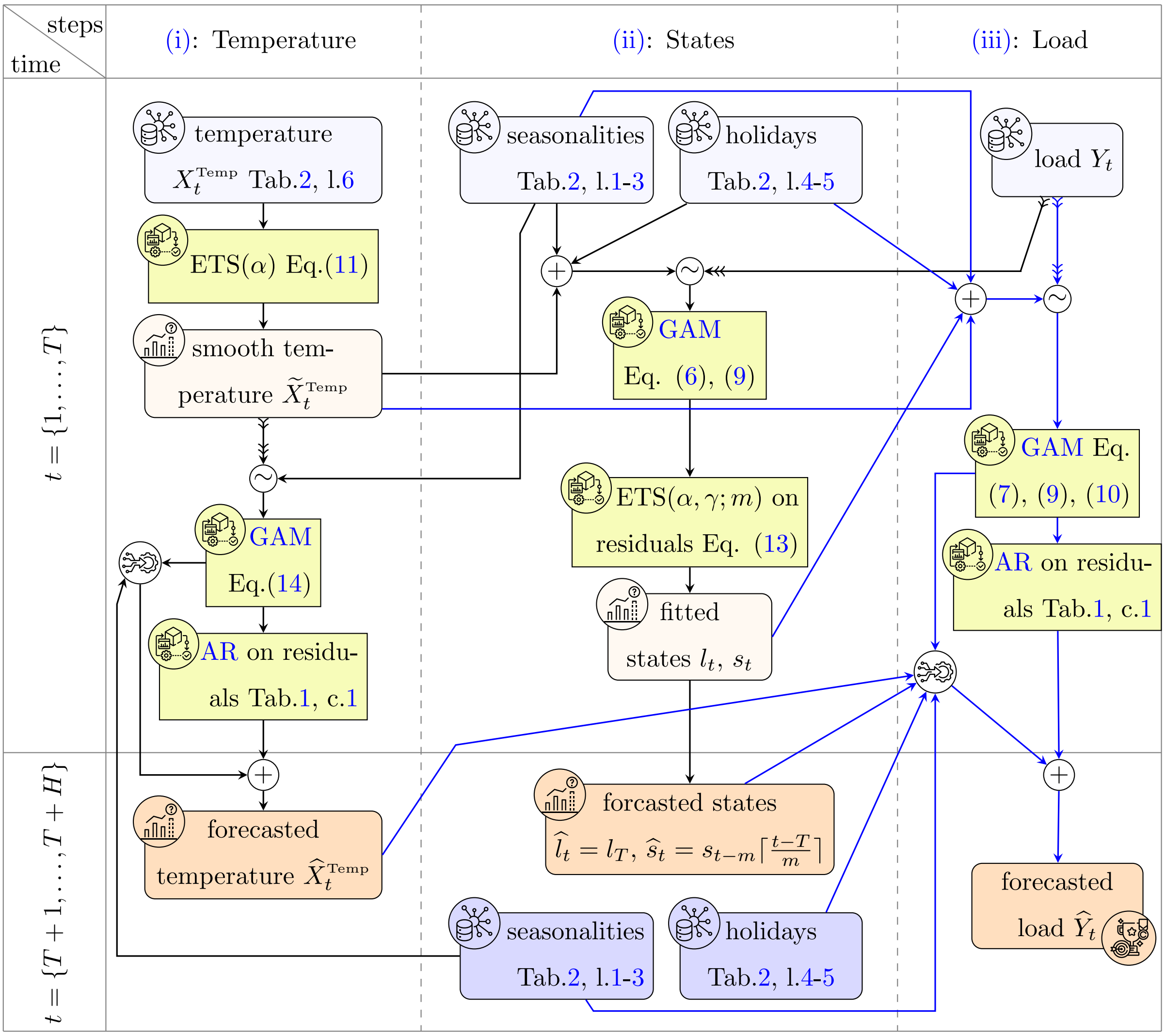

Figure 6: Flowchart illustrating the GAM-based load forecasting model. External inputs are framed by blue boxes, model outputs by orange boxes and models by yellow boxes. Lighter shades represent in-sample times and darker shades out-of-sample times. Blue arrows connect inputs and models of the final load forecast.

### *4.2. Data Preprocessing*

Utilizing temperature data $\text{Temp}_t \in \mathbb{R}$ observed at weather stations and time information $t \in 1, ..., T$, variables $X_t^m \in \mathbb{R}$ of our GAM model are defined in Table 2 and described in the subsequent. As outlined in Section 2, our dataset encompasses hourly observations, resulting in $S = 24$ observations per day. Nevertheless, the methodology introduced in the subsequent sections is formulated for arbitrary observations per day $S \in \mathbb{N}_{\geq 3}$, allowing the analysis of data collected at different resolutions. For instance, it can be applied to commonly observed half-hourly data for $S = 48$ or quarter-hourly data for $S = 96$. Without loss of generality, we assume that the time series begins on January 1st at $0:00$.

The variable $X_t^{\text{HoY}}$ is designed to ensure a consistent representation of the time of the year across leap and non-leap years. This is guaranteed by a day of the year variable $\text{DoY}_t$ adjusted for leap years. Specifically, the variable $\text{DoY}_t \in \{0, ..., 364\}$ counts the day of the year starting with zero on January 1st and ending with 364 in non-leap years. If in a leap year, February 28th and 29th are assigned the same value of $\text{DoY}_t$ and subsequent values of $\text{DoY}_t$ have the same value as in non-leap years.

The variable $X_{t_j}^{\text{HldP}}$ has discrete values in the set $\{0, 1, ..., h_{\text{hldp}}\}$ representing the count of time steps from the beginning to the end of the holiday period. It is active only during the holiday period taking the value zero otherwise. In the proposed model the holiday period starts on $t_{\text{hldp}}$ set to December 18th at $0:00$ and ends on January 6th at $23:00$ covering $t_{\text{hldp}} = 20S$ hours.

Analogously to $X_t^{\text{HldP}}$, the variables of a weekday and fixed date holiday $X_t^{\text{HldW}}$ and $X_t^{\text{HldF}}$ have discrete values in the set $\{0, 1, ..., 2b + S\}$ representing the holiday bridged to the previous and subsequent day by counting its time steps and being zero otherwise. Specifically, $b = 12$ is chosen and holidays are bridged from midday of the preceding day to midday of the following day. Since holiday information is usually available only as the date of the holiday, preprocessing of this date to $t_{\text{hld}}$ is necessary. In both GAM models, the smooth term in $X_t^{\text{HldF}}$ will additionally be multiplied by an impact factor $X_t^{\text{Impact}}$ to account for the variation in the load reduction regarding the day of the week on which the holiday occurs.

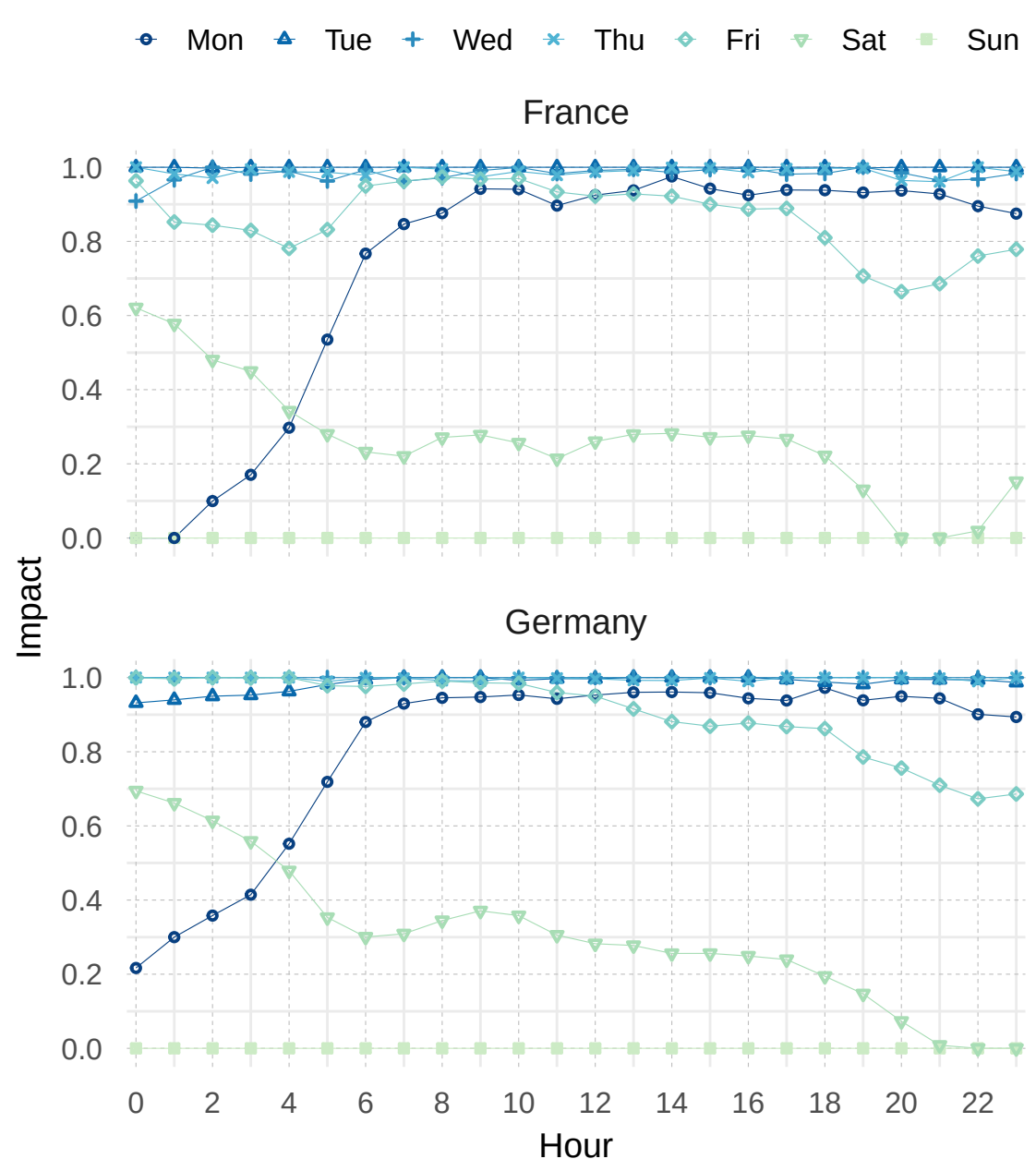

Figure 7: Evaluation of $X_t^{\text{Impact}}$ for French and German hourly load data from January 1st, 2015 to December 31st, 2018.

This impact factor, illustrated in Figure 7, is defined by the values of $I_t$ bounded to $[0, 1]$. The numerator of $I_t$ calculates the deviation of the weekly load profile, which is expressed as the median load for the hour $X_t^{\text{HoW}} \in \{0, ..., 7S - 1\}$ of the week in the data, from the Sunday load profile, which is defined as the median load for the corresponding hour of the day on a Sunday $X_t^{\text{HoD}} + 6S \in \{6S, ..., 7S - 1\}$ in the data. The denominator of $I_t$ calculates the deviation from the peak weekly load profile, which is defined as the mean of the median loads for the corresponding hour of the day on a Tuesday $X_t^{\text{HoD}} + S \in \{S, ..., 2S - 1\}$, on a Wednesday $X_t^{\text{HoD}} + 2S \in \{2S, ..., 3S - 1\}$ and on a Thursday $X_t^{\text{HoD}} + 3S \in \{3S, ..., 4S - 1\}$ from the Sunday profile. Thus, the impact factor $X_t^{\text{Impact}}$ provides an hourly ratio of the weekly load profile at time $t$ to the peak weekly load profile scaled with respect to the Sunday load profile, which has an impact factor of zero. Consequently, the multiplication of a smooth term in the fixed date holiday variable $X_t^{\text{HldF}}$ with the impact variable $X_t^{\text{Impact}} = X_{t \bmod 7S}^{\text{Impact}}$ serves to eliminate the load effect if the holiday falls on a Sunday. Conversely, the impact is nearly maximal when the holiday falls on a Tuesday, Wednesday or Thursday. If the holiday falls on a Monday, Friday or Saturday, the impact is diminished.

The temperature variable $X_t^{\text{Temp}}$ is calculated as the mean of temperatures $\text{Temp}_t$ in $n_{\text{Temp}} = 5$ most populated cities[5] per country.

[5]Note that by this selection criterion, the cities used to calculate the average temperature may change over the study period. Additionally, note that various approaches exist to aggregate air temperatures per country. However, research by Neumann et al. (2023) and Sobhani et al. (2019) found no significant improvement over simple averaging compared to more

To preserve seasonal periodicities and account for a quick adjustment of human behavior to the local time, we adjust for daylight saving time changes in all variables except the meteorological temperature. Specifically, for all deterministic variables, clock change is considered non-existing. For the observed load data the missing hour in March is linearly interpolated and the double hour in October is averaged. The temperature variable $X_t^{\text{Temp}}$ is defined in UTC, but adjusted in the same way, after ETS-smoothing and forecasting by the seasonal GAM.

Single outliers in the load data are handled via a lasso-based linear regression of load against a dummy for each observation, an Hour of the Day dummy and the running median of load for three consecutive observations. From the estimated $\lambda$-grid a model is chosen by squared BIC and the load time series is adjusted by the estimated observation dummies. Due to regression against the running median of three observations, only single outliers, but not consecutive outliers are detected. By including dummies for the hour of the day, the method takes the interday variation into account.

Table 2: Definition of variables $X_t^m \in \mathbb{R}$ in the proposed two-stage GAM load forecasting model.

| **Variable** | **Definition** | **Codomain** |
|---|---|---|
| Hour of day (HoD) | $X_t^{\text{HoD}} = t \bmod S$ | $\{0, ..., S-1\}$ |
| Hour of week (HoW) | $X_t^{\text{HoW}} = t - t^* \bmod 7S$, where $t^*$ is Monday at 0:00 in $\{1, ..., 7S\}$. | $\{0, ..., 7S-1\}$ |
| Hour of year (HoY) | $X_t^{\text{HoY}} = \text{DoY}_t \times S + X_t^{\text{HoD}}$ | $\{0, ..., 365S-1\}$ |
| Holiday period (HldP) | $X_t^{\text{HldP}} = \begin{cases} X_{t+1}^{\text{HoY}} - X_{t_{\text{hldp}}}^{\text{HoY}} & \text{for } t \in \{t_{\text{hldp}}, \dots, t_{\text{hldp}} + h_{\text{hldp}} - 1\} \\ 0 & \text{otherwise,} \end{cases}$ where $t_{\text{hldp}} \in \{1, ..., T\}$ is the first occurrence of the holiday period of length $h_{\text{hldp}}$. | $\{0, 1, ..., h_{\text{hldp}}\}$ |
| Weekday and fixed date holiday (HldW, HldF) | $X_t^{\text{Hol}} = \begin{cases} X_{t+1}^{\text{HoY}} - X_{t_{\text{hld}}-b}^{\text{HoY}} & \text{for } t \in \{t_{\text{hld}} - b, ..., t_{\text{hld}} + b + S - 1\}, \\ 0 & \text{otherwise,} \end{cases}$ for hol $\in$ {HldW,HldF}, where $t_{\text{hld}} \in \{1, ..., T\}$ is the first occurrence of the weekday or fixed date holiday and $b \in \{0, ..., S\}$ bridges the holiday to the previous and subsequent day by $b$ time steps. | $\{0, 1, ..., 2b + S\}$ |
| Impact | $X_t^{\text{Impact}} = \min(\max(0, I_t, 1))$ <br> $I_t = \frac{\text{med}(Y_{t^*} \mid X_{t^*}^{\text{HoW}} = X_t^{\text{HoW}}) - \text{med}(Y_{t^*} \mid X_{t^*}^{\text{HoW}} = X_t^{\text{HoD}} + 6S)}{\text{mean}(\text{med}(Y_{t^*} \mid X_{t^*}^{\text{HoW}} = X_t^{\text{HoD}} + iS)_{i=1}^3) - \text{med}(Y_{t^*} \mid X_{t^*}^{\text{HoW}} = X_t^{\text{HoD}} + 6S)}$ | $[0, 1]$ |
| Temperature (Temp) | $X_t^{\text{Temp}} = \frac{1}{n_{\text{Temp}}} \sum_{i=1}^{n_{\text{Temp}}} \text{Temp}_t^i$ | $\mathbb{R}$ |

sophisticated methods.

*4.3. Model Specification*

Our load forecasting model is composed of two stages of GAM models in sums of smooth terms $\mu_1$ and $\mu_1 + \mu_2$, respectively, and an AR process on the second-stage residuals $\mathcal{E}_t$:

$$Y_t = \upsilon + \mu_1(\boldsymbol{X}_t) + \varepsilon_t, \tag{6}$$

$$Y_t = o + \underbrace{\mu_1(\boldsymbol{X}_t) + \mu_2(\boldsymbol{X}_t)}_{\text{smooth effects in the covariates}} + \underbrace{\mathcal{E}_t}_{\text{autoregressive effects}}, \tag{7}$$

$$\mathcal{E}_t = \phi_1 \mathcal{E}_{t-1} + \ldots + \phi_p \mathcal{E}_{t-p} + e_t, \tag{8}$$

where $\upsilon, o \in \mathbb{R}$.

The additive terms $\mu_1$ and $\mu_2$ in equations (6) and (7) are defined as

$$\begin{aligned}\mu_1(\boldsymbol{X}_t) = {} & \underbrace{f^{\text{Temp},1}(\widetilde{X}_t^{\text{Temp},1}) + f^{\text{Temp},2}(\widetilde{X}_t^{\text{Temp},2})}_{\text{temperature effects}} \\ & + \underbrace{f^{\text{HoD}}(X_t^{\text{HoD}}) + f^{\text{HoW}}(X_t^{\text{HoW}}) + f^{\text{HoY}}(X_t^{\text{HoY}})}_{\text{seasonal effects}} \\ & + \underbrace{f^{\text{HoY,HoD}}(X_t^{\text{HoY}}, X_t^{\text{HoD}}) + f^{\text{HoY,HoW}}(X_t^{\text{HoY}}, X_t^{\text{HoW}})}_{\text{seasonal interaction effects}} \\ & + \underbrace{f^{\text{HldW}}(X_t^{\text{HldW}}) + X_t^{\text{Impact}} f^{\text{HldF}}(X_t^{\text{HldF}}) + f^{\text{HldP}}(X_t^{\text{HldP}})}_{\text{holiday effects}},\end{aligned} \tag{9}$$

$$\mu_2(\boldsymbol{X}_t) = \underbrace{X_t^{\text{Level}} f^{\text{Level}}(X_t^{\text{HoD}}) + X_t^{\text{Season}} f^{\text{Season}}(X_t^{\text{HoD}})}_{\text{mid-term states effects}} \tag{10}$$

and $\widetilde{X}_t^{\text{Temp}_{1,2}}$ are the fitted values of

$$X_t^{\text{Temp}_{1,2}} \sim \text{ETS}(\alpha_{1,2}) \tag{11}$$

for $\alpha_1 = 1/S$ and $\alpha_2 = 1/14S$. All smooth effects $f^m$ in the covariates are constructed as degree three, thus cubic, P-splines.

The mid-term states, i.e. level and seasonal, covariates $X_t^{\text{Level}}$, $X_t^{\text{Season}}$ in (10) are derived from the residual term $\varepsilon_t$ of the first stage GAM model (6), which captures temperature, seasonal and holiday effects. Since the model (6) does not capture a unit root, this information remains in $\varepsilon_t$.

An intuitive approach to extract the unit-root states from $\varepsilon_t$ would be to estimate an ETS model on $\varepsilon_t$ with the multiple prevailing seasonalities, i.e. daily, weekly and annual seasonalities. However, an hourly ETS model with multiple seasonalities and a decent lead time would result in unfeasibly high computational expenses.

To receive a computationally feasible option, we apply temporal aggregation such that the frequency of $\varepsilon_t$ is reduced to a weekly interval. This approach decreases the computational costs by reducing the data frequency by a factor of $7S$ and by eliminating the daily and weekly seasonality in ETS estimation. Nonetheless, it retains the typical unit root behavior in load levels and the annual seasonality.

The temporal aggregation of $\varepsilon_t$ is defined as its average across specific hours of the week given by the set $C$:

$$r_\tau = \frac{1}{|C|} \sum_{c \in C} \varepsilon_{7S(\tau-1)+c} \tag{12}$$

Specifically, we choose $C = \bigcup_{n=0}^{4}\{8,...,19\} + nS$ representing the hours from 8:00 to 19:00 for each working day of the week (Mon-Fri). This choice aligns with the peak load definition commonly used in European electricity markets for trading electricity futures.

An ETS model with annual seasonality is applied to this averaged noise (see Tab. 1, col. 3):

$$r_\tau \sim \text{ETS}(\alpha, \gamma; m = 52) \tag{13}$$

with smoothing parameters $\alpha$ and $\gamma$ for level and seasonal component, respectively.

The level and seasonal covariate $X_t^{\text{Level}}$ and, $X_t^{\text{Season}}$ for all $t \in \{1, ..., T\}$ result from linear interpolation of the level and seasonal component $l_\tau$ and $s_\tau$ of (13) with extended boundaries. Thus, $X_t^{\text{Level}} = l_{\tau_1}$ and $X_t^{\text{Level}} = l_{\tau_T}$ for $t < \min(C)$ and $t > \max(C) + 7S\tau_T$, repectively, and analogously for $X_t^{\text{Season}}$.

The information loss from temporal aggregation in (12), which leads to diminished daily and weekly variation of covariates $X_t^{\text{Level}}$ and $X_t^{\text{Season}}$ is mitigated by multiplication[6] of both covariates with cyclic smooth terms in $X_t^{\text{HoD}}$. While theoretically, it would be preferable to multiply with a smooth term in $X_t^{\text{HoW}}$ the computational complexity renders it impractical. Instead, a smooth term in $X_t^{\text{HoD}}$ suffices to capture the relevant information.

For all covariates with periodic structure, thus for seasonal and holiday covariates, spline types are cyclic, see Table 3. By appropriate positioning of the first and last knot, this guarantees under penalization of parameter deviation that $f^m(X_{np_m}^m) \approx f^m(X_{(n+1)p_m-1}^m)$ for period length $p_m$ of the covariate $X^m$, $m \in \{\text{HoD}, \text{HoW}, \text{HoY}, \text{HldW}, \text{HldF}, \text{HldP}\}$ and analogously $f^m(X_{np_{m_1}}^{m_1}, X_{np_{m_2}}^{m_2}) \approx f^m(X_{(n+1)p_{m_1}-1}^{m_1}, X_{(n+1)p_{m_2}-1}^{m_2})$ for period lengths $p_{m_{\{1,2\}}}$ of the covariates $X^{m_{\{1,2\}}}$, $(m_1, m_2) \in \{(\text{HoD}, \text{HoY}), (\text{HoW}, \text{HoY})\}$. Thus, while $X_{np_m}^m \ll X_{(n+1)p_m-1}^m$, e.g. $X_{(n+1)p_m-1} - X_{np_m} = 365S$ for $m = \text{HoY}$, the transition from one period of $f^m$ to the subsequent is approximately seamless, as required by the application. By way of illustration, it would be unsuitable for the smooth term in the HoY variable to be disconnected at the end of the year.

In line with the discrete range of values of seasonal and holiday covariates their intermittent knots[7] are set to discrete. The chosen hyperparameters for the smooth terms, i.e. the basis ranks, knots and spline type are listed in Table 3.

To emphasize, the novelties of our GAM modeling approach are:

(i) The individual handling of effects induced by each holiday separately as a smooth term in a discrete

[6]We intentionally choose a multiplication with a smooth term in $X_t^{\text{HoD}}$ instead of the full interaction, thus a two-dimensional smooth term in covariates $X_t^{\text{HoD}}$ and $X_t^{\text{Level}}$ or $X_t^{\text{Season}}$. This choice aims to prevent an additional time-varying effect besides the level and seasonal states effect on load. The additive effect of level and seasonal states on load was retrieved from the ETS model in (13), where the gradual variation of these effects is governed by the estimated smoothing parameters $\alpha$ and $\gamma$. Embedding the derived covariates $X_t^{\text{Level}}$ and $X_t^{\text{Season}}$ into smooth terms would allow for these effects to additionally vary according to the estimated smooth term. Consequently, an extra time-varying effect on load would be introduced potentially overlaying and thus hindering interpreting the effect of unit root states.

[7]Note that the number of knots is determined by the basis dimension $k_m$. More precisely, the number of knots is $k_m + 1$ for cyclic splines. Moreover, recall that knots are spaced equidistant for the traditionally defined P-Splines. Consequently, the basis dimension $k_m$ has to be a devisor of the period length.

Table 3: Definition of covariates $X_t^m \in \mathbb{R}$ in the proposed two-stage GAM load forecasting model. For the non-cyclic smooth terms in temperatures, knots were not specified but set according to the common method on knot spacing, see Wood (2006), within the range of the covariate.

| **Smooth Term** | **Basis Dimension** | **Knots** | **Spline Type** |
|---|---|---|---|
| $f^{\text{Temp},\{1,2\}}(\widetilde{X}_t^{\text{Temp},\{1,2\}})$ | $k_{\text{Temp}} = 6$ | | non-cyclic |
| $f^{\text{Temp}}(X_t^{\text{HoD}})$ | $k_{\text{HoD}} = S$ | $\{0, ..., S\}$ | cyclic |
| $f^{\text{HoW}}(X_t^{\text{HoW}})$ | $k_{\text{HoW}} = 7S$ | $\{0, ..., 7S\}$ | cyclic |
| $f^{\text{HoY}}(X_t^{\text{HoY}})$ | $k_{\text{HoY}} = 12$ | $\{0, 365S/12, 365S/6, ..., 365S\}$ | cyclic |
| $f^{\text{HoY,HoD}}(X_t^{\text{HoY}}, X_t^{\text{HoD}})$ | $(k_{\text{HoY}}, k_{\text{HoD}}) = (12, S/3)$ | $\{0, 365S/12, 365S/6, ..., 365S\}\times$ $\{0, 3, 6...S\}$ | cyclic |
| $f^{\text{HoY,HoW}}(X_t^{\text{HoY}}, X_t^{\text{HoW}})$ | $(k_{\text{HoY}}, k_{\text{HoD}}) = (12, 7)$ | $\{0, 365S/12, 365S/6, ..., 365S\}\times$ $\{0, S, 2S, ..., 7S\}$ | cyclic |
| $f^{\text{HldW}}(X_t^{\text{HldW}})$ | $k_{\text{HldW}} = S/3$ | $\{0, 6, 12, ..., 2S\}$ | cyclic |
| $f^{\text{HldF}}(X_t^{\text{HldF}})$ | $k_{\text{HldF}} = S/3$ | $\{0, 6, 12, ..., 2S\}$ | cyclic |
| $f^{\text{HldP}}(X_t^{\text{HldP}})$ | $k_{\text{HldP}} = S/3$ | $\{0, 60, 120, ..., 20S\}$ | cyclic |
| $f^{\text{Level}}(X_t^{\text{HoD}})$ | $k_{\text{Level}} = S/3$ | $\{0, 3, 6, ..., S\}$ | cyclic |
| $f^{\text{Season}}(X_t^{\text{HoD}})$ | $k_{\text{Season}} = S/3$ | $\{0, 3, 6, ..., S\}$ | cyclic |

activation[8] variable. By sensibly specifying this activation variable, additionally, the load reduction on previous and subsequent days induced by holidays that occur each year on a Tuesday or Thursday, e.g. Ascension Day or Corpus Christi in Germany, are captured. This approach is applied similarly to the extended holiday period of the Christmas vacation time. Moreover, we dampen the influence of holidays with a fixed date depending on the weekday they occur. With this individual holiday modeling, we differ from conventional methods that model holidays by a single holiday or weekend dummy, see Ziel (2018), assigning all holidays the same effect on load.

(ii) We leverage a P-Spline GAM to forecast mid-term ETS-smoothed typical temperature patterns by seasonalities. This approach addresses the data scarcity, regarding accurate weather and satellite-imagery information, for forecasting periods ranging from several weeks to a year.

(iii) To account for the mid-term forecasting horizon and the entailing gradual change of equilibrium load due to economic or socioeconomic variations, we incorporate state components. These components are obtained by an ETS model on the residuals of the first-stage GAM and thus exhibit a persistent trend behavior.

### *4.4. Estimation and Forecasting*

Both GAM models (see Eq. (6) and (7)) are estimated under discretization by fast REML, according to (3), where in all splines parameter deviation of degree two is penalized. Model coefficients of the first-stage GAM (see Eq. (6)) are passed as initial parameters to the second-stage GAM (see Eq. (7)) to speed up computation time. We allow for shrinking smooth terms to zero[9] and reduce effective sample size to obtain smoother fits, in a BIC-like range, see Wood et al. (2015). The AR process $\mathcal{E}_t \sim \text{AR}_{\text{AIC}}(\phi; p_{\max})$ (see Eq. (8)) is estimated as specified in Subsection 3.2 with $p_{\max} = 8 \times 7S$. For estimation of the ETS model (see Eq. (13)) the trace MAE (see Eq. (5)) for a horizon $h_{\text{ETS}}$ of two weeks is minimized.

[8] To guarantee that the non-active and active hours of these holidays are transferred to the smooth splines, their identifiability constraints are set to $f^m(0) = 0$.

For non-deterministic variables, i.e. for the smooth temperature, level and seasonal variables, forecasts for each step of the horizon $H$ have to be obtained.

The smooth temperature forecasts $\widehat{X}_t^{\text{Temp}\{1,2\}}$, $t \in \{T+1, ..., T+H\}$ result from, firstly, estimating the GAM model

$$\begin{aligned} \widetilde{X}_t^{\text{Temp}j} &= c^{\text{Temp}} + \sum_{i=1}^{k_{\text{HoD},1}} \beta_i^{\text{HoD}} b_i^{\text{HoD}}(X_t^{\text{HoD}}) + \sum_{i=1}^{k_{\text{HoY}}} \beta_i^{\text{HoY}} b_i^{\text{HoY}}(X_t^{\text{HoY}}) \\ &+ \sum_{j=1}^{k_{\text{HoY}}} \sum_{i=1}^{k_{\text{HoD},2}} \beta_{ji} b_i^{\text{HoD}}(X_t^{\text{HoD}}) b_j^{\text{HoY}}(X_t^{\text{HoY}}) + e_t \end{aligned} \tag{14}$$

with cyclic cubic P-Splines, basis dimensions $k_{\text{HoD}} \in \{S, S/3\}$, $k_{\text{HoY}} = 6$ and corresponding knots $\{0, ..., S\}$, $\{0, 365S/6, ..., 365S\}$, $\{0, 365S/6, ..., 365S\} \times \{0, 3, 6...S\}$ by fast REML with degree two parameter deviation penalties (see Eq. (3)). Secondly, an AR-process is estimated to the residuals of (14) as specified in Subsection 3.2:

$$\widehat{e}_t \sim \text{AR}_{\text{AIC}}(\phi; p_{\max}) \tag{15}$$

with $p_{\max} = 4 \times 7S$ .

The level and seasonal states forecasts result by $\widehat{l}_t = l_T$ and $\widehat{s}_t = s_{t-m\lceil \frac{t-T}{m} \rceil}, t > T$ from the estimated ETS model (13).

### *4.5. Benchmark Models*

Ten benchmark models, alongside the proposed GAM model, will be jointly analyzed in a forecasting study described in Section 5.

As baseline for comparison, two seasonal random walk models with weekly and annual periods (**SRWW**, **SRWA**) that involve no parameter estimation are evaluated:

$$\widehat{Y}_t = Y_{t-m\lceil \frac{t-T}{m} \rceil}, t \in \{T+1, ..., T+H\}, m \in \{7S, 365S\}$$

Furthermore, three variations of the Vanilla Model due to Hong (2010); Hong et al. (2014) are analysed: a deterministic variant with no temperature information (**VanDet**), a basic variant with actual temperature information (**VanBas**) and a variant with additional recent temperature information (**VanRec**,

[9]If this is not specified in the estimation process, smooth terms are shrunk to the null space of the penalty functional $\mathcal{P}(.)$ (see Eq. (4)), i.e. the subspace of splines $f(x) = \sum_{i=1}^{k} \beta_i b_i(x)$ such that $\mathcal{P}(f) = 0$. For cyclic P-splines, this is, regardless of the polynomial and penalty degree, the one-dimensional space of constant polynomials. In the non-cyclic case, this is for the penalty degree $p$, regardless of the polynomial degree, the p-dimensional subspace of $(p-1)$-degree polynomials in the basis span.

(16)):

$$Y_t = \beta_0 + \beta_1 X_t^{\text{Trend}} + \boldsymbol{\beta}_2 \boldsymbol{X}_t^{\text{Month}} + \boldsymbol{\beta}_3 \boldsymbol{X}_t^{\text{Week}} + \boldsymbol{\beta}_4 \boldsymbol{X}_t^{\text{Hour}} + \boldsymbol{\beta}_5 \boldsymbol{X}_t^{\text{Week}} \boldsymbol{X}_t^{\text{Hour}} + p(X_t^{\text{Temp}}) + p(\widetilde{X}_t^{\text{Temp}_1}) + p(\widetilde{X}_t^{\text{Temp}_2}) + \epsilon_t, \tag{16}$$

where $X_t^{\text{Trend}} = t$ represents a linear trend, $\boldsymbol{X}_t^{\text{Month}}$, $\boldsymbol{X}_t^{\text{Week}}$, $\boldsymbol{X}_t^{\text{Hour}}$ are class variables for the month, weekday and hour at time $t \in \{1, ..., T+H\}$, $\widetilde{X}_t^{\text{Temp}_i}$ are the fitted values of (11) and

$$p(x) = \beta_6 x + \beta_7 x^2 + \beta_8 x^3 + \boldsymbol{X}_t^{\text{Month}}(\beta_9 x + \beta_{10} x^2 + \beta_{11} x^3) + \boldsymbol{X}_t^{\text{Hour}}(\beta_{12} x + \beta_{13} x^2 + \beta_{14} x^3). \tag{17}$$

Modelling equations of **VanDet** and **VanBas** are obtained from (16) by $p(X_t^{\text{Temp}}) = p(\widetilde{X}_t^{\text{Temp}_1}) = p(\widetilde{X}_t^{\text{Temp}_2}) = 0$ and $p(\widetilde{X}_t^{\text{Temp}_1}) = p(\widetilde{X}_t^{\text{Temp}_2}) = 0$, respectively. The temperature forecasts $\widehat{X}_t^{\text{Temp}_{1,2}}$, $t \in \{T+1, ..., T+H\}$, see (14), from our proposed model are applied.

The linear models **VanDet**, **VanBas**, **VanRec** are selected as benchmarks due to their prevalence in load forecasting, see e.g.,GEFCom 2012 and 2017 Davis et al. (2014); Hong et al. (2019) and Høverstad et al. (2015); Ziel & Liu (2016); Luo et al. (2018); Xie et al. (2018), their low computational cost and their interpretability, which makes them a potential competitor for the proposed GAM model in its key advantageous attributes.

An Autoregressive Weekly Difference Model (**ARWD**) based on Haben et al. (2019), is chosen as a benchmark since it has been among the most accurate methods in short-term load forecasting in Ziel (2018):

$$Y_t - \mu_t \sim \text{AR}_{\text{AIC}}(\boldsymbol{\phi}; p_{\max}),$$

where $\mu_t$ is the sample mean of load for each hour of the week and the AR process is estimated as specified in Subsection 3.2.

A combined ETS and Seasonal Random Walk Model with Seasonal and Trend decomposition using Loess (**STL**) from the forecast package, see Hyndman & Athanasopoulos (2018), is applied as benchmark. Daily, weekly and annual seasonal periods are applied for the decomposition and forecasts are computed by seasonal random walks for the seasonal components and an AIC selected ETS model for the seasonally adjusted time series.

As additional non-linear model, a Feed Forward Neural Network (**FNN**) with one hidden layer and linear output, which resulted as best performing model in Bashiri Behmiri et al. (2023) is evaluated:

$$Y_t = \upsilon + \sum_{j=1}^{k_{\text{FNN}}} v_j g\left(\sum_{j=1}^{m} w_{ij} \boldsymbol{X}_t\right),$$

where $\boldsymbol{X}_t = (\widetilde{X}_t^{\text{Temp}_1}, \widetilde{X}_t^{\text{Temp}_2}, X_t^{\text{HoD}}, X_t^{\text{HoW}}, X_t^{\text{HoY}}, X_t^{\text{HldP}}, X_t^{\text{HldW}}, X_t^{\text{HldF}})$ for $\widetilde{X}_t^{\text{Temp}_i}$ the fitted values of (11), $m = |\boldsymbol{X}_t|$, $k_{\text{FNN}} = m/2 + 1$ and sigmoid activation function $g(x) = 1/(1 + \exp^{-x})$.

Lastly, to evaluate the skill of the GAM-based temperature forecasts (see Eq. (14)-(15)), we compare them with climatology-based approaches (**CLIM**, **CL-ETS**) as inputs for our load forecasting model, as proposed by Bashiri Behmiri et al. (2023). Both approaches calculate the forecast as average of the corresponding in-sample temperatures, with **CL-ETS** applying ETS-smoothing (see Eq. (11)) before averaging:

$$\widehat{X}_t^{\text{Temp}} = 1/|\mathcal{D}| \sum_{d \in \mathcal{D}} X_d^{\text{Temp}}, \quad \widehat{\widetilde{X}}_t^{\text{Temp}\{1,2\}} = 1/|\mathcal{D}| \sum_{d \in \mathcal{D}} \widetilde{X}_d^{\text{Temp}\{1,2\}},$$

for $\mathcal{D} = \{X_t^{\text{HoY}} + 1 + (i365S)_{i=0}^{\lfloor T/365S \rfloor}\}$, $t \in \{T+1, \dots, T+H\}$.

We designed an assortment of benchmarks with models covering the most relevant features for mid-term load forecasting, i.e. daily, weekly and annual, autoregressive and temperature effects, that includes both common linear and non-linear modeling techniques, and encompasses frequently applied benchmarks in load forecasting.

Given the heterogeneous dataset including countries with significant electric heating and cooling, we considered it essential to include benchmark models capturing temperature effects. We do not include a benchmark model capturing holiday effects, since the impact of these effects on forecasting accuracy is assessed in Section 6 by a comparison of our GAM model with and without holiday effects. For a comprehensive forecasting study on holiday handling in load forecasting, we refer to Ziel (2018).

## 5. Forecasting Study and Evaluation Design

To evaluate the proposed GAM model we conducted a rolling window forecasting study on the benchmark models, see Subsection 4.5, and our GAM model successively leaving out one modeling component considered a novelty, i.e. the smooth effects in the temperature (**Temp**), the mid-term smooth effects in the level and seasonal component (**States**) and, lastly, the smooth effects in all three categories of holidays (**Hld**). Additionally, the performance of the GAM model leaving out the autoregressive post-processing (see Eq. 10) is assessed (**AR**). For the study, more than 9 years of load data, from January 2015 to February 2024 were used. In-sample data in each forecasting experiment comprised 4 years of data. Recall that data are observed at hourly resolution, thus in-sample data comprises $4 \times 365 \times 24$ observations per experiment. With each forecasting experiment the window of in-sample observations is rolled forward by 24 hours, whereby the first hour to forecast is always 9 a.m. The forecasting horizon is chosen as 52 weeks, i.e. $H = 168 \times 52$. Due to the high overlap of hourly forecasts for a horizon of 52 weeks, the forecasting experiments of a study are highly correlated. Consequently, $N = 100$ experiments are randomly sampled from the total number of approximately $1460 = 4 \times 365$ forecasting experiments. This corresponds to approximately one experiment every two weeks.

We evaluate forecasting accuracy using the mean absolute error MAE[10] in three distinct ways: Firstly, to gain a comprehensive understanding, $\text{MAE}(\widehat{\mathcal{E}}_h) = 1/N \sum_{n=1}^{N} |Y_{T+h} - \widehat{Y}_{T+h,n}|$ is averaged over all hours in the forecasting horizon $h = 1, ..., H$. Secondly, the significance of a difference in the overall average MAE between two models is tested via the Diebold-Mariano (DM) test introduced by Diebold & Mariano (1994). Thirdly, to identify temporal variations, $\text{MAE}(\widehat{\mathcal{E}}_h)$ is averaged over the weeks $\{1, ..., 52\}$ in the one-year forecasting horizons $\{1, ..., H\}$ and over the hours $\{1, ..., 24\}$ in $\{1, ..., H\}$.

[10] As a reminder, a regression-type GAM model is designed to forecast the mean, making Root Mean Square Error (RMSE) the appropriate error measure. However, MAE, optimal for median forecasts, is applied, since it is more robust and due to its intuitive interpretability common in this type of literature. For a comparison of overall performance, RMSE is considered additionally in Table 8 in Appendix B. We specifically do not consider mean absolute percentage error (MAPE), though also common in this type of literature, due to its widely acknowledged theoretical shortcomings, see Goodwin & Lawton (1999); Fildes (1992); Hyndman & Athanasopoulos (2018); Franses (2016). For more details on error measures, we refer to Gneiting (2009).

## 6. Results and Discussion

### 6.1. Overall Performance

Regarding overall performance, see Table 4, our proposed GAM models (**GAM**, **\AR**, **\HLD**, **\Temp**, **\States**) distinctly outperform all benchmark models except the **FNN** and GAM-type models (**CLIM**, **CL-ETS**) in every country included in the study. Our best-performing GAM model has superior performance compared to all benchmarks across all countries.

Notably, **SRWW** and all three vanilla benchmark models (**VanDet**, **VanBas**, **VanRec**) exhibit low forecasting accuracy. Similar deficiencies in forecasting accuracy of Vanilla benchmarks were observed by e.g. Browell & Fasiolo (2021) or Ziel & Liu (2016); Gaillard et al. (2016) in GEFCom2014, see Hong et al. (2016), and Ziel (2019) in GEFCom2017, see Hong et al. (2019). However, including actual temperatures and recency temperature effects, even when only forecasted by seasonality, improves Vanilla forecasting accuracy.

When incorporated in GAM models, temperature effects also improve forecasting accuracy. Specifically, we find the following performance hierarchy in most countries: **\Temp** $<$ **CLIM** $<$ **CL-ETS** $<$ **GAM** which proves the skill of our temperature model over a climatological approach.

Among the non-GAM-based benchmark models, **FNN** performs best, followed by the simple **SRWA** and the **STL** model.

Table 4: Color-coded (red: low, blue: high) forecasting accuracy in terms of $\overline{\text{MAE}(\widehat{\mathcal{E}}_h)}$ for $h = 1, \ldots, 168 \times 52$ in GW with the best model per country bolded.

| | SRWW | SRWA | STL | VanDet | VanBas | VanRec | ARWD | FNN | CLIM | CL-ETS | GAM | \AR | \HLD | \Temp | \States | DA TSO |
|---|---|---|---|---|---|---|---|---|---|---|---|---|---|---|---|---|
| Austria | 0.865 | 0.448 | 0.494 | 1.071 | 1.010 | 1.005 | 0.627 | 0.369 | 0.386 | 0.382 | 0.377 | 0.378 | 0.410 | 0.401 | **0.336** | 0.332 |
| Belgium | 0.958 | 0.559 | 0.580 | 1.058 | 1.000 | 0.987 | 0.726 | 0.498 | 0.471 | 0.466 | 0.460 | 0.461 | 0.485 | 0.484 | **0.435** | – |
| Bulgaria | 0.767 | 0.347 | 0.386 | 0.638 | 0.521 | 0.509 | 0.570 | 0.283 | 0.388 | 0.304 | 0.286 | 0.286 | 0.294 | 0.300 | **0.279** | 0.134 |
| Czech Rep. | 1.037 | 0.432 | 0.477 | 1.043 | 0.933 | 0.920 | 0.745 | 0.346 | 0.361 | 0.353 | 0.341 | 0.342 | 0.391 | 0.380 | **0.315** | 0.119 |
| Denmark | 0.447 | 0.249 | 0.244 | 0.582 | 0.551 | 0.538 | 0.336 | 0.240 | 0.205 | 0.203 | **0.200** | 0.201 | 0.209 | 0.207 | 0.217 | 0.049 |
| Estonia | 0.157 | 0.067 | 0.073 | 0.150 | 0.124 | 0.123 | 0.118 | 0.055 | 0.057 | 0.057 | 0.056 | 0.056 | 0.057 | 0.060 | **0.053** | 0.070 |
| Finland | 1.555 | 0.732 | 0.683 | 1.172 | 0.711 | 0.720 | 1.208 | 0.580 | 0.545 | 0.560 | 0.542 | 0.543 | 0.553 | 0.578 | **0.535** | 0.169 |
| France | 11.499 | 4.536 | 4.839 | 8.193 | 5.723 | 5.603 | 8.750 | 3.744 | 4.009 | 3.908 | 3.761 | 3.782 | 3.804 | 4.438 | **3.527** | 0.948 |
| Germany | 5.635 | 3.339 | 3.336 | 7.978 | 7.786 | 7.684 | 4.160 | 2.766 | 2.394 | 2.341 | **2.324** | 2.340 | 2.693 | 2.439 | 2.404 | 2.089 |
| Greece | 0.968 | 0.567 | 0.615 | 0.953 | 0.865 | 0.834 | 0.690 | 0.462 | 0.482 | 0.473 | 0.457 | 0.461 | 0.472 | 0.502 | **0.452** | 0.174 |
| Hungary | 0.564 | 0.326 | 0.369 | 0.605 | 0.574 | 0.564 | 0.407 | 0.274 | 0.289 | 0.300 | 0.296 | 0.297 | 0.317 | 0.298 | **0.252** | 0.225 |
| Italy | 3.984 | 2.440 | 3.129 | 6.303 | 6.231 | 6.129 | 2.658 | 2.118 | 2.272 | 2.183 | 2.159 | 2.176 | 2.452 | 2.373 | **1.854** | 0.674 |
| Latvia | 0.085 | 0.050 | 0.050 | 0.128 | 0.120 | 0.119 | 0.067 | 0.045 | 0.036 | 0.037 | 0.036 | **0.036** | 0.039 | 0.038 | 0.039 | 0.020 |
| Lithuania | 0.154 | 0.087 | 0.094 | 0.205 | 0.193 | 0.190 | 0.113 | 0.073 | 0.077 | 0.077 | 0.076 | 0.076 | 0.081 | 0.078 | **0.068** | 0.037 |
| Montenegro | 0.067 | 0.042 | 0.041 | 0.068 | 0.063 | 0.058 | 0.051 | 0.039 | 0.038 | 0.037 | 0.036 | 0.036 | **0.036** | 0.041 | 0.037 | 0.024 |
| Netherlands | 1.428 | 0.844 | 1.252 | 1.610 | 1.535 | 1.515 | 1.120 | 0.987 | 0.845 | 0.850 | 0.844 | 0.846 | 0.844 | **0.816** | 0.927 | 1.259 |
| Poland | 1.884 | 1.083 | 1.187 | 2.688 | 2.633 | 2.604 | 1.289 | 0.809 | 0.837 | 0.807 | 0.792 | 0.795 | 0.957 | 0.840 | **0.722** | 0.532 |
| Portugal | 0.570 | 0.325 | 0.425 | 0.854 | 0.856 | 0.835 | 0.390 | 0.285 | 0.310 | 0.308 | 0.305 | 0.308 | 0.326 | 0.314 | **0.246** | 0.158 |
| Romania | 0.726 | 0.469 | 0.483 | 0.840 | 0.786 | 0.784 | 0.551 | 0.387 | 0.386 | 0.383 | 0.377 | 0.380 | 0.403 | 0.400 | **0.349** | 0.096 |
| Serbia | 0.694 | 0.352 | 0.367 | 0.617 | 0.498 | 0.503 | 0.584 | 0.391 | 0.313 | 0.308 | **0.299** | 0.302 | 0.305 | 0.331 | 0.345 | 0.141 |
| Slovakia | 0.364 | 0.247 | 0.208 | 0.403 | 0.374 | 0.375 | 0.279 | 0.201 | 0.180 | 0.175 | **0.172** | 0.173 | 0.188 | 0.185 | 0.179 | 0.076 |
| Slovenia | 0.202 | 0.118 | 0.121 | 0.241 | 0.232 | 0.230 | 0.147 | 0.104 | 0.098 | 0.097 | **0.096** | 0.096 | 0.105 | 0.100 | 0.097 | 0.046 |
| Spain | 2.752 | 1.688 | 2.059 | 3.682 | 3.634 | 3.544 | 2.004 | 1.596 | 1.620 | 1.589 | 1.546 | 1.564 | 1.665 | 1.608 | **1.487** | 0.309 |
| Sweden | 3.289 | 1.169 | 1.237 | 2.507 | 1.526 | 1.521 | 2.491 | 0.931 | 0.951 | 0.965 | 0.931 | 0.934 | 0.949 | 0.971 | **0.913** | 0.404 |
| Sum | 40.651 | 20.519 | 22.748 | 43.586 | 38.480 | 37.893 | 30.081 | 17.582 | 17.552 | 17.165 | 16.768 | 16.870 | 18.034 | 18.183 | **16.071** | 8.085 |

With performance in the range of the GAM models, **FNN** is a considerable competitor for mid-term load forecasting. However, while good at capturing the effect of the multifaceted characteristics of load it lacks interpretability[11] in these. From our GAM model we will infer an intuitive and detailed understanding of how different factors influence load across specific time frames in Subsection 6.6.

As detailed in Table 6, the **FNN** is also disadvantageous to our GAM models in terms of computation time. The computation time of the **FNN** could be decreased by providing appropriate inputs. Consequently, the proposed GAM model may be employed as a first step in a model ensemble with a FNN. With the effect of most characteristics, in particular, the non-stationarity addressed, a FNN can focus on remaining non-linearities in a rigorously reduced time. This, additionally, mitigates the risk of suboptimal solutions in FNN optimization. By way of illustration, model ensembling with a machine learning model on the residuals of a GAM resulted in a 5% gain in model performance in the study of Nedellec et al. (2014) and is advised in De Vilmarest & Goude (2022).

The good performance of the simple **SRWA** benchmark highlights the importance of accounting for annual seasonality in mid-term load forecasting. The proposed GAM model incorporates this by a smooth yearly seasonal term and the ETS-modeled seasonal component. Additionally, the model could potentially be improved by decomposing with regard to the yearly seasonal component as a first modeling step and modeling it separately by a SRWA or by incorporating yearly lags in addition to the current lags limited by the boundary $p_{\max} = 8 \times 168$.

Since a direct comparison with the day-ahead TSO[12] forecasts is not reasonable due to the different forecasting horizons, they are not colored according to the red-green color scheme. However, the low difference between our best-performing GAM model to the TSO forecasts (e.g. in Austria, Estonia, Germany, Hungary, Netherlands and Poland the DA TSO outperforms by less than 70%) highlights the good performance of our model and its applicability in the industry, most importantly for mid-term production and maintenance planning.

Within our GAM models, **\States** achieves the lowest average MAE in most countries. In the second-highest number of countries, the **GAM** model incorporating all effects (autoregressive, holidays, temperature and states) has the lowest average MAE.

We observe a similar picture of overall performance for the entire European market assessed based on the sum of errors across all considered countries, see Table 4 last line, as well as for the RMSE as metric, see Table 8 in Appendix A.

### *6.2. Performance during COVID-19*

Figure 8, depicting the error measure for each forecasting experiment in France and Germany, reveals that high overall errors in the **GAM** vs. the **\States** model originate from the springtime in 2020. During this time the load level suddenly and considerably decreased due to COVID-19 lockdowns, as confirmed by studies such as Guo et al. (2024); Obst et al. (2021); Bahmanyar et al. (2020); Narajewski

[11] While with FNN models one can interpret the sensitivity of load to changes in input variables, see e.g. Behm et al. (2020), they fail to provide insight into how the model's input factors contribute to the assembly of load over the forecasted time frame, as demonstrated in Figures 14 and 18.

[12] TSO day-ahead forecasts are included in Table 4 for countries with less than 1% of forecasts missing.

& Ziel (2020). This phenomenon is also evident in load levels represented by the in-sample residuals of the first-stage GAM (see Eq. (6)) and the 52-week moving average load[13] depicted in Figure 9 from May 24th, 2016 to May 18th, 2020 and 2021, respectively.

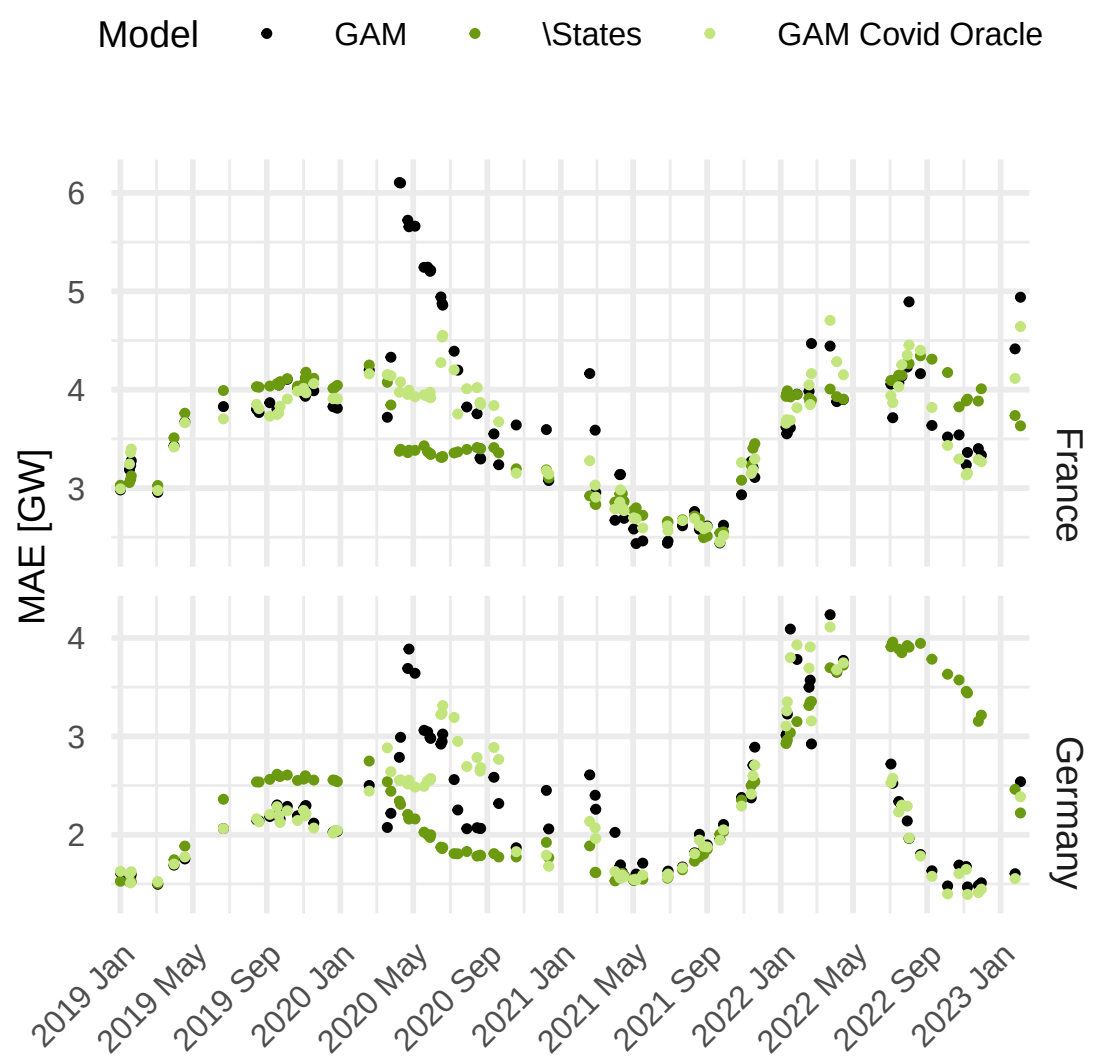

Figure 8: MAE($\widehat{\mathcal{E}_n}$)=$1/H \sum_{h=1}^{H} |Y_{T+h} - \widehat{Y}_{T+h,n}|$ in GW for $n = 1, ..., N$.

Note that the first lockdowns in Germany (March 22nd to May 4th, 2020) and France (March 17th to May 10th, 2020), see ACAPS COVID-19 Government Measures Dataset ACAPS (2024), had ended prior to the in-sample residuals in Figure 9. Hence, we observe from Figure 9 that load levels in terms of first-stage GAM residuals during these lockdowns not only decreased sharply but also rebounded quickly, almost reaching pre-lockdown levels in France, after restrictions had been lifted. This pattern of a sharp load reduction followed by rapid recovery after initial lockdowns aligns with findings from other studies, including Guo et al. (2024); Buechler et al. (2022).

Thus, contrary to the ETS model assumption of a persistent level, which equals the last fitted value (see Tab. 1, col. 3), in the forecasting horizon, load remained at low levels only temporarily for forecasting experiments during the springtime in 2020. As a result, the constant ETS load level forecast, see the green line in Figure 9, remained unrealistically low for much of the one-year forecasting horizon. This underestimation is also evident when comparing the 52-week moving average load with the estimated and forecasted ETS level in the same figure. Incorporating this constant level forecast into the GAM model (see Eq. (7) and (10)) led to accumulating errors over the forecast horizon, resulting in the observed error difference between the **GAM** and **\States** models shown in Figure 8.

To capture the non-persistent structural break in the load data caused by COVID-19 lockdowns, we extended the ETS model for the aggregated residuals $r_\tau$ (see Eq. (13)) of the first-stage GAM model. Specifically, we incorporated a COVID-lockdown dummy variable $X_\tau^{\text{Covid}}$ as an explanatory variable. This dummy satisfies $X_\tau^{\text{Covid}} = 1$ if at least one day of the week fell within a lockdown period and $X_\tau^{\text{Covid}} = 0$ otherwise. Given that $X_\tau^{\text{Covid}}$ can only be determined retrospectively using current knowledge of lockdown dates, we refer to this model as **GAM Covid Oracle**[14]:

$$r_\tau = l_{\tau-1} + s_{\tau-m} + a_{\text{Covid}} X_\tau^{\text{Covid}} + \epsilon_\tau, \text{ for } l_\tau = l_{\tau-1} + \alpha\epsilon_\tau,\ s_\tau = s_{\tau-m} + \gamma\epsilon_\tau,\ m = 52 \tag{18}$$

[13]To facilitate comparison with other load values, the annual moving average load in Figure 9 has been centered by subtracting its mean.

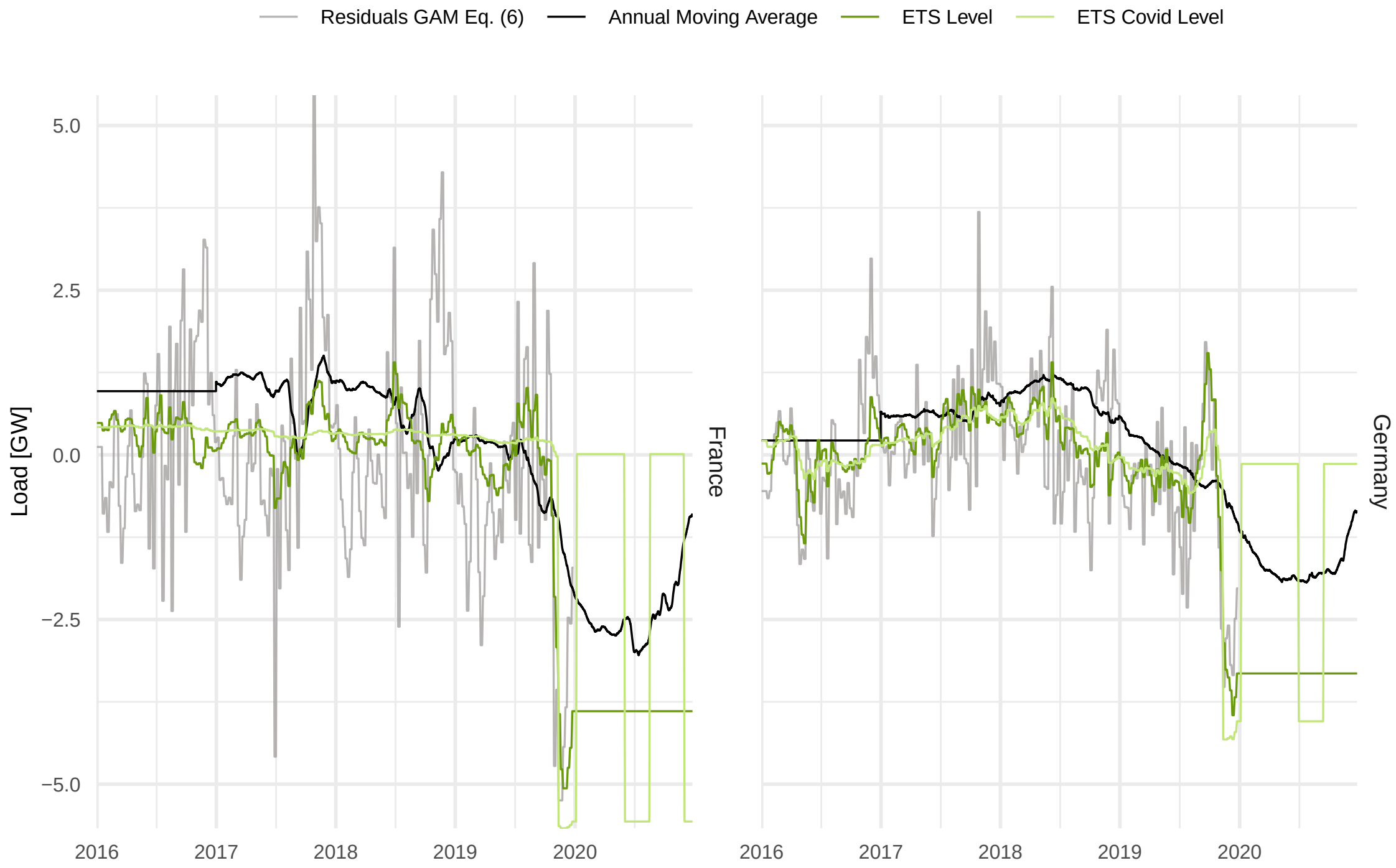

Figure 9: Yearly, i.e. 52 weeks, moving average load (black), estimated and forecasted level for models **GAM** (dark green) and **GAM Covid Oracle** (light green), and in-sample residuals (grey) of the first stage GAM (Eq. (6)) from May 24th, 2016 to May 18th, 2021.

For a detailed explanation of explanatory variables in ETS models and the stability and forecastability of such models, we refer to Svetunkov (2023). The level covariate $X_t^{\text{Level}}$ for $t \in \{1, ..., T+H\}$ is derived through linear interpolation with extended boundaries of the estimate $l_\tau + \widehat{a}_{\text{Covid}} X_\tau^{\text{Covid}}$ and forecast $\widehat{l}_{\tau_T+h} + \widehat{a}_{\text{Covid}} X_{\tau_T+h}^{\text{Covid}}$ with $\widehat{l}_{\tau_T+h} = \widehat{l}_{\tau_T}$ for all $h = 1, \ldots, H$. The seasonal covariate $X_t^{\text{Season}}$ is calculated as outlined in Subsection 4.3. Data on COVID-19 lockdown dates for France and Germany were sourced from the ACAPS COVID-19 Government Measures Dataset ACAPS (2024).

Figure 9 reveals that the **GAM Covid Oracle** model forecasts a level in the pre-COVID range and a 5.6 GW and 3.3 GW decrease in load for France and Germany, respectively, during lockdowns. As shown in Figure 8, this reduces modeling errors for experiments during the springtime of 2020 resulting in the lowest average MAE and RMSE (Table 5) in Germany. In France, though averaged errors have reduced when COVID-19 information is integrated, the **\States** model maintains the lowest average errors.

Despite retrospective access to COVID-19 lockdown dates, accurately capturing these structural breaks within an ETS framework remains challenging. As highlighted by Svetunkov (2023); Koehler et al. (2012), ETS models struggle with forecasting and parameter estimation when faced with data discontinuities. Moreover, Gurriaran et al. (2023) emphasize the general difficulty of modeling the sudden break caused by COVID-19 lockdowns, even when incorporating additional data on restrictions, such as school and workplace closures, travel bans and mobility restrictions, which are in part unavailable for European countries.

[14]Since the **GAM Covid Oracle** model relies on retrospective data, it is excluded from comparisons in the other subsections of Section 6.

Table 5: Color-coded (red: low, blue: high) forecasting accuracy in terms of $\overline{\text{MAE}(\widehat{\mathcal{E}}_h)}$ and $\overline{\text{RMSE}(\widehat{\mathcal{E}}_h)}$ (in brackets) for $h = 1, \dots, 168 \times 52$ in GW with the best model per country bolded.

| | GAM | \States | GAM Covid Oracle |
|---|---|---|---|
| France | 3.761 ( 5.037) | **3.527** ( **4.779**) | 3.601 ( 4.853) |
| Germany | 2.324 ( 3.007) | 2.404 ( 3.071) | **2.292** ( **2.978**) |

To effectively address these exceptional periods De Vilmarest & Goude (2022); Obst et al. (2021) suggest incorporating GAM-based models into adaptive frameworks, which were developed further to a probabilistic forecasting model by De Vilmarest et al. (2024). However, while De Vilmarest & Goude (2022) report that MAEs could nearly be halved by adapting GAM parameters to different states using a Kalman filter, their focus on short-term forecasting and a single COVID-19 wave limits the direct applicability of their approach to our year-ahead forecasting context, which encompasses multiple lockdown periods with evolving human behavior.

Beyond the COVID-19 impact, load is influenced by a complex interplay of numerous macroeconomic, socioeconomic and energy-economic factors, as detailed in Subsection 2.1. The presence of these numerous, partially unquantifiable factors renders modeling the socio-economic load component challenging, as highlighted by Goude et al. (2014). Due to the absence of comprehensive exogenous data on these factors across the 24 European countries considered in this study, applying ETS models with additional exogenous variables beyond COVID-19 lockdowns to improve the modeling of the socio-economic component is impractical.

Instead, to address the limitation of a persistent socio-economic forecast inherent in the ETS framework, time-varying ETS models, as proposed by Koehler et al. (2012), or ensembles with models that forecast states converging to zero, such as stationary autoregressive models, should be explored.

Lastly, we argue against dropping unit-root states from the model, i.e. against the **\States** model, since the presence of a unit root in the load time series is indicated by the estimated smoothing parameters $\alpha$ and $\gamma$ (see Eq. (13)). Across the forecasting experiments, the average values of $\alpha$ and $\gamma$ ($\overline{\alpha} = 0.33$, $\overline{\gamma} = 0.022$ in France, $\overline{\alpha} = 0.31$, $\overline{\gamma} = 0.021$ in Germany) are different from zero. As the ETS model nests a non-unit root model for smoothing parameters approximately zero, these findings strongly suggest the presence of a unit root in the load level and a weak unit root in the annual seasonal component. Moreover, previous studies on short-term load forecasting demonstrated the benefits of incorporating non-stationary trends, see e.g. Pierrot & Goude (2011); Goude et al. (2014); Nedellec et al. (2014); Ziel (2019).

### *6.3. DM-Test Results*

Regarding the significance of the findings described in Subsection 6.1 for France and Germany, we observe from Figure 10 that including autoregressive, holiday and temperature effects improves accuracy significantly in both countries. Including mid-term state effects improved accuracy significantly only in Germany, whereas for the French load data, the GAM model including all but the effect of the state has significantly better forecasting accuracy compared to all other models.

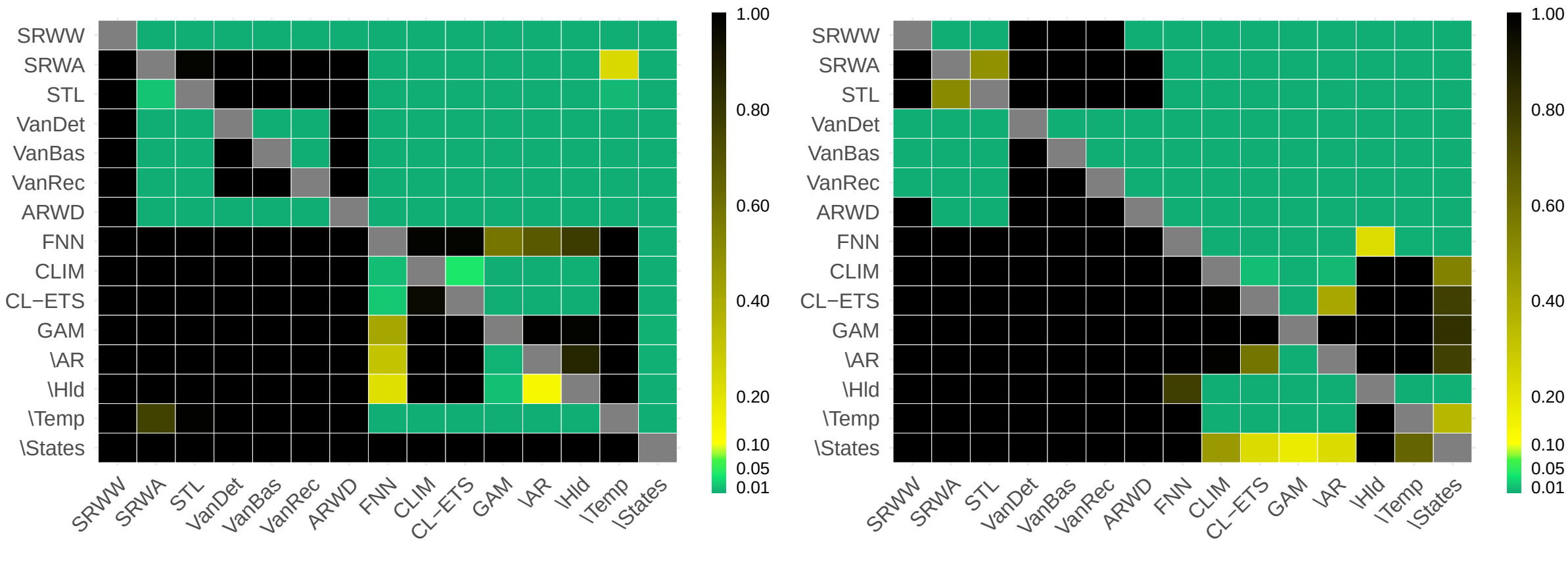

(a) France

(b) Germany

Figure 10: P-values of the DM-test for France and Germany.

For comparison, in the study by Bashiri Behmiri et al. (2023) on Italian load forecasts including temperature effects forecasted by random walk and climatology models only showed significantly improved RMSE in a DM-test for short-term horizons. Specifically, Bashiri Behmiri et al. (2023) forecast short-term (up to 7 days) daily temperatures by the last known daily averaged temperature and mid-term daily temperatures by the average daily temperature of the previous five years.

The divergent outcomes may be attributed to a different methodology in terms of integrating and forecasting temperature. Whereas they integrate actual temperature as a linear spline with one predefined knot in linear regression models comparable to **VanBas**, we incorporate ETS-smoothed temperatures and model their effect using non-linear smooth terms. In our study, **\Temp** is significantly outperformed by **CLIM**, which is, in turn, significantly outperformed by **CL-ETS**, highlighting the advantage of incorporating climatology-based temperature effects as non-linear smooth terms and the added benefit of ETS-smoothing in the in-sample period to account for delayed temperature effects on load. Furthermore, **CL-ETS** is significantly outperformed by **GAM**, underscoring the benefits of smooth temperature forecasts that incorporate short-term autoregressive information from climatology-based forecasts.

Additionally, their methodology lacks a daily variation, as they forecast daily load from daily forecasted temperatures. In contrast, the non-linear relation between load and temperature varies during the day as we see e.g. from Figure 5 and was observed by Fan & Hyndman (2012). When true temperatures were incorporated instead of forecasts, they observed a significant enhancement in accuracy, ranging from 10% to 20%. These values align closely with our findings (see Subsection 6.4).

Comparing the GAM-type models to benchmarks, except **FNN**, we fail to reject the null hypothesis of outperformance only for **\Temp**, which is surpassed by **SRWA** in France. This underscores the significance of integrating temperature effects in a country where electricity demand is influenced by heating and cooling.

When comparing the GAM-type models to the **FNN** benchmark, we observe a significantly better performance in France only for the **\States**, while in Germany, all GAM-type models, except for **\HLD**, exhibit significantly better performance. In the study by Bashiri Behmiri et al. (2023), the similarly structured FNN model also outperformed regression-based models. However, the difference in RSME was significant only for a short-term horizon of up to one week. The diminished performance of the regression-based models for the short-term horizon may be caused by the linearity of these models lacking variation in load throughout the week. Specifically, they use linear model structures that explain the weekly seasonality with dummies for Saturday, Sunday and Monday.

### *6.4. Weekly and Hourly Performance*

For deeper insights, how MAE changes throughout the one-year forecasting horizons and the hours in one day, Figure 11 shows the weekly and hourly averaged MAEs for the three best-performing non-GAM-based benchmarks alongside our GAM-type models for France and Germany.

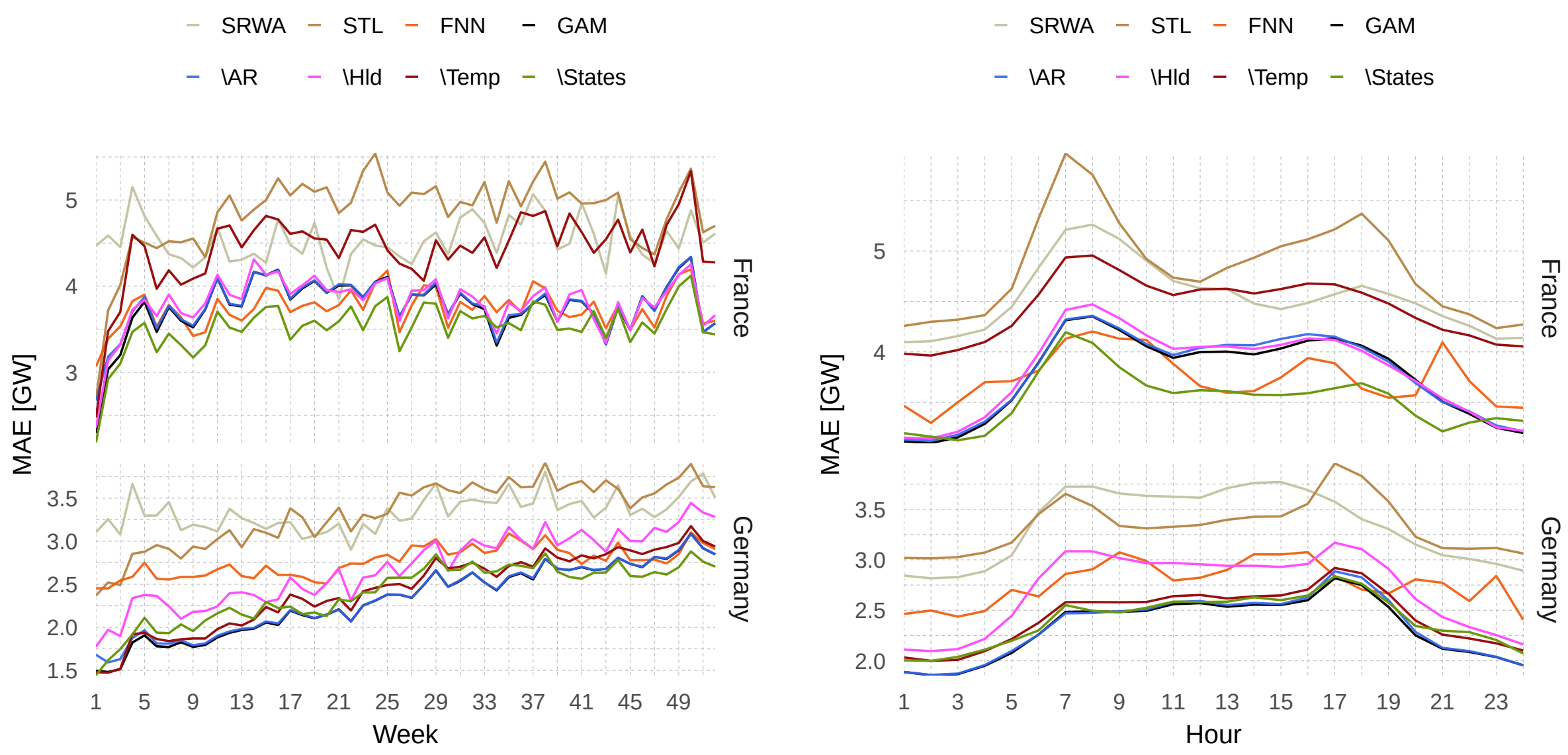

(a) $\overline{\text{MAE}(\widehat{\mathcal{E}}_h)}$ in GW for $h = h \bmod 52$ (b) $\overline{\text{MAE}(\widehat{\mathcal{E}}_h)}$ in GW for $h = h \bmod 24$

Figure 11: Mean $\text{MAE}(\widehat{\mathcal{E}}_h)$ over weeks of the year and hours of the day for France and Germany.

We observe that models with autoregressive components (**STL** and GAM-type models) exhibit low initial errors that increase gradually as the forecast horizon extends. This effect is particularly visible in Germany. In France, a steep increase in errors within the first two weeks diminishes the visibility of the gradual increase in the mid-term. In comparison, the **SRWA** benchmark displays consistently high and middle-ranging errors across the horizon. This emphasizes the importance of autoregressive components for the short horizons and in France a capturing of short-term weather effects by autoregressive components.

Comparing GAM-type models for a short-term horizon (2-3 weeks) we observe in Germany that the full **GAM** model and **\Temp** model perform best (approx. 10%, 15% and 7% better than **\AR**, **\HLD** and **\States**). In France, in particular, incorporating temperature effects contributes to improved forecasting accuracy (approx. 10% for horizons greater than a few days). This, along with the worse performance of the **STL** benchmark (approx. 15% and 60% in France and Germany, respectively) during this time frame highlights that capturing the multifaceted characteristics of load time series, and specifically factors affecting load on the mid-term, e.g. mid-term states components and holiday effects, improves performance not only in mid-term but also for short-term forecasting. Consequently, the proposed GAM-type models are also a suitable choice for short-term load forecasting. Additionally, in both countries, the short-term performance (2-3 weeks) of our **GAM** model is enhanced by autoregressive postprocessing, with the **GAM** model showing an approximate 10% improvement over **\AR**.

The importance of the load characteristics varies between France and Germany. In France, we observe from comparing the different GAM-type models for the entire forecasting horizon that the incorporation of temperature results in the greatest improvement in forecasting accuracy (approx. 20%), whereas including holiday effects has no observable effect. Conversely, in Germany, the inclusion of nuanced holiday smoothing terms brings the greatest improvement (approx. 10%), a value consistent with the findings of the comprehensive study conducted by Ziel (2018) on holiday handling in German load forecasting. In Germany, temperature effects only provide a minor enhancement (approx. 5%). This again highlights that particularly for countries with temperature-sensitive load profiles, including temperature information is beneficial even when accurate temperature forecasts are unavailable and the temperature model relies solely on seasonal and autoregressive information.

Though our GAM-based temperature model relying solely on seasonalities, autoregressive information and four years of historical data effectively predicts typical smoothed temperature patterns, it cannot capture the effects of extreme weather events and long-term climatic changes. Since such meteorological uncertainties are inherently stochastic drivers of temperatures, they can only be adequately addressed within a probabilistic forecasting model. Our proposed temperature model can be straightforwardly expanded into a probabilistic framework by incorporating a more extensive historical dataset and a trend component to capture potential climate change impacts. Probabilistic temperature forecasts can then be generated iteratively by sampling errors from a normal distribution, with parameters estimated from the in-sample residuals of the second-stage AR model. By using a longer historical dataset that captures a broader range of meteorological conditions, including extreme events, the model has increased variability in in-sample residuals. This, in turn, increases the variability of forecasted temperature and load distributions, making the model more suitable for risk assessment.

To enhance the effectiveness of our holiday handling in France, considering the introduction of a second holiday period could prove advantageous. As noted by Pierrot & Goude (2011), French load data typically exhibit a significant decline, particularly on weekdays, during August.

In order to achieve greater accuracy improvements through the inclusion of temperature effects in Germany, it may be beneficial to explore the most influential weather stations using GAM's GCV score, as proposed by Dordonnat et al. (2016). By this, regional disparities in electric heating and cooling patterns across Germany could be incorporated.

Regarding the change of performance for the different hours of the day, see Figure 11, we observe higher errors during peak demand hours (7-9 a.m., 5-7 p.m.) and lower errors during night and midday hours of low demand for all models (approx. 25%, 15% for GAM-type models in France, Germany, respectively). Consequently, in further studies, multivariate models for high and low-demand hours, or in its most finely resolved form, for 24 hours as they are advocated in GAM load models by e.g. Pierrot & Goude (2011); Nedellec et al. (2014); Dordonnat et al. (2016); Gaillard et al. (2016), should be analyzed.

For the **FNN** benchmark, this error pattern for high- and low-demand hours is not clearly visible. In comparison to GAM-type models, **FNN** errors are particularly high for the early night hours and, in Germany, for the early morning hours. Again, this draws our attention to the potential of a model ensembling our proposed **GAM** model and a FNN. While the **GAM** model outperforms in capturing interday temporal dependencies with interpretable smooth terms in the variable $X^{\text{HoD}}$, a second-step FNN can address remaining non-linearities.

### *6.5. Computation Time*

Computation times for our GAM models and non-GAM-based benchmarks on a standard laptop are listed for France and Germany in Table 6. Additionally, Table 7 [15] lists the computation times of each GAM-type model for the estimation of each submodel and forecasting.

While all benchmarks, except **FNN**, considered in this study performed significantly worse than the proposed GAM model, most of their computational times, of under half a second are remarkably fast due to their simple linear structure. Note that a considerable amount of their computation time goes into outlier preprocessing, see Subsection 4.2. The computation time of the GAM models remains within a reasonable range, varying between 2.6 and 4.8 seconds. The computation time of the **FNN** benchmark is substantially higher when compared to the best-performing GAM model. However, when compared to more sophisticated machine learning models in load forecasting, our **FNN** still has low computation times. Other machine learning approaches for mid-term load forecasting e.g. deep neural networks used by Han et al. (2019) and model ensembles of machine learning approaches by Li et al. (2023) or Agrawal et al. (2018) took approximately half an hour to one hour for training.

Table 6: Mean computation time of models for $4 \times 365 \times 24$ in-sample observations and $365 \times 24$ ahead forecasting on a standard laptop for France and Germany.

| | **SRWW** | **SRWA** | **STL** | **VanDet** | **VanBas** | **VanRec** | **ARWD** | **FNN** | **GAM** | **\AR** | **\HLD** | **\Temp** | **\States** |
|---|---|---|---|---|---|---|---|---|---|---|---|---|---|
| **Time (FR)** | 0.07 | 0.07 | 1.12 | 0.08 | 0.08 | 0.09 | 0.47 | 6.73 | 4.27 | 3.22 | 3.09 | 3.40 | 2.59 |
| **Time (DE)** | 0.06 | 0.06 | 1.10 | 0.07 | 0.07 | 0.08 | 0.38 | 9.08 | 4.84 | 3.41 | 3.04 | 4.19 | 2.88 |

Table 7 shows that autoregressive postprocessing increases computational costs due to AR model estimation (see Eq. (8)) and additional forecasting time by approximately one second in both countries. Despite this overhead, the significant improvement in forecasting accuracy (see Subsection 6.3), and its necessity to ensure applicability of our model for horizons of 2-3 weeks (see Subsection 6.4), justify the inclusion.

Comparing all GAM-type models, the greatest computational difference in total estimation time lies between **GAM** and **\States**, as the latter omits ETS model and second-stage GAM estimation. However, the presence of a unit root in the load time series, supported by empirical evidence (see Subsection 6.2) and theoretical considerations (see Subsection 2.1), necessitates the inclusion of the state component.

[15] Note that the difference in the computation times between Table 6 and 7 is attributed to the additional steps of outlier handling, data preparation, and assembling required for model estimation. The sum of computation times listed in Table 7 pertain exclusively to the estimation of the submodels.

Table 7: Mean computation time for estimation of submodels and forecasting for $4 \times 365 \times 24$ in-sample observations and $365 \times 24$ ahead forecasting on a standard laptop for France and Germany.

| **Time** | **FR** | **GAM** | **\AR** | **\HLD** | **\Temp** | **\States** | **DE** | **GAM** | **\AR** | **\HLD** | **\Temp** | **\States** |
|---|---|---|---|---|---|---|---|---|---|---|---|---|
| GAM Eq. (6) | | 1.05 | 1.05 | 0.53 | 0.51 | 1.05 | | 1.17 | 1.17 | 0.47 | 0.70 | 1.17 |
| ETS (13) | | 0.05 | 0.05 | 0.05 | 0.04 | | | 0.05 | 0.05 | 0.05 | 0.05 | |
| GAM Eq. (7) | | 1.38 | 1.36 | 0.70 | 0.81 | | | 1.60 | 1.44 | 0.53 | 1.41 | |
| AR Eq. (8) | | 0.84 | | 0.90 | 1.09 | 0.91 | | 1.08 | | 1.09 | 1.09 | 1.09 |
| Forecasting | | 0.07 | 0.03 | 0.05 | 0.07 | 0.07 | | 0.07 | 0.03 | 0.06 | 0.07 | 0.07 |
| Sum | | 3.39 | 2.49 | 2.23 | 2.52 | 2.03 | | 3.97 | 2.69 | 2.19 | 3.32 | 2.33 |

Computation times for the temperature model (see **\Temp**) are excluded from Tables 6 and 7 as it is a component used by multiple benchmarks and not directly part of the load GAM models. Average computation times for the temperature model are 1.02 and 1.00 seconds in France and Germany, respectively.

### *6.6. **GAM** Model Interpretation*

To complement our analysis, Figure 12 demonstrates the holiday-specific load-reducing effect in France and Germany. The figure depicts the estimated smooth effects in (7) for weekday holidays (Good Friday, Easter Monday) and fixed day holidays (Labour Day, All Saints Day) fitted to the first in-sample data window of the forecasting study. Notably, the impact of Good Friday in France, a regional holiday celebrated by only approximately 5% of the population, is negligible. On the other hand, All Saints' Day, a regional holiday celebrated by roughly 57% of the German population, exhibits a weaker effect compared to non-regional holidays, but still a more pronounced effect than Good Friday in France. The remaining non-regional holidays have a similar load-reducing effect. Load reduction is minimal on the preceding day, peaks around midday of the holiday and gradually declines to negligible levels by the following day. A similar load reduction profile is observed for most other holidays, which are depicted for France and Germany in Figures 17-16 in the Appendix B. From these Figures, we observe that smooth terms in further holidays celebrated by a negligible percentage of the population, e.g. Abolition of Slavery (FR), Abolition de l'esclavage (FR)[16], St. Stephen's Day (FR), International Women's Day (DE) World Children's Day (DE), and Repentance and Prayer Day (DE), are shrunk mostly to zero, comparable to Good Friday in France. Note that the specification of the estimation method, see Subsection 4.4, allows for shrinking smooth terms to zero, instead of the null space. For our nuanced holiday modeling, this is particularly suitable and makes a pre-preprocessing of holiday data with an impact factor, according to their population share, obsolete.

In addition to the effect induced by specific holidays, Figures 13 and 14 depict the effect induced by all load characteristics considered in our GAM model (see Eq. (7)). The figures show the estimated smooth terms in holidays and the holiday period, seasonalities including interactions, temperature and mid-term states component fitted to the first in-sample data window of the forecasting study in the Christmas holiday time in France and the Easter holiday time in Germany, respectively. Furthermore, the fitted and actual loads reduced by the estimated intercept are pictured.

[16] Abolition of Slavery and Abolition de l'esclavage, though having the same name, are different holidays in French overseas territories.

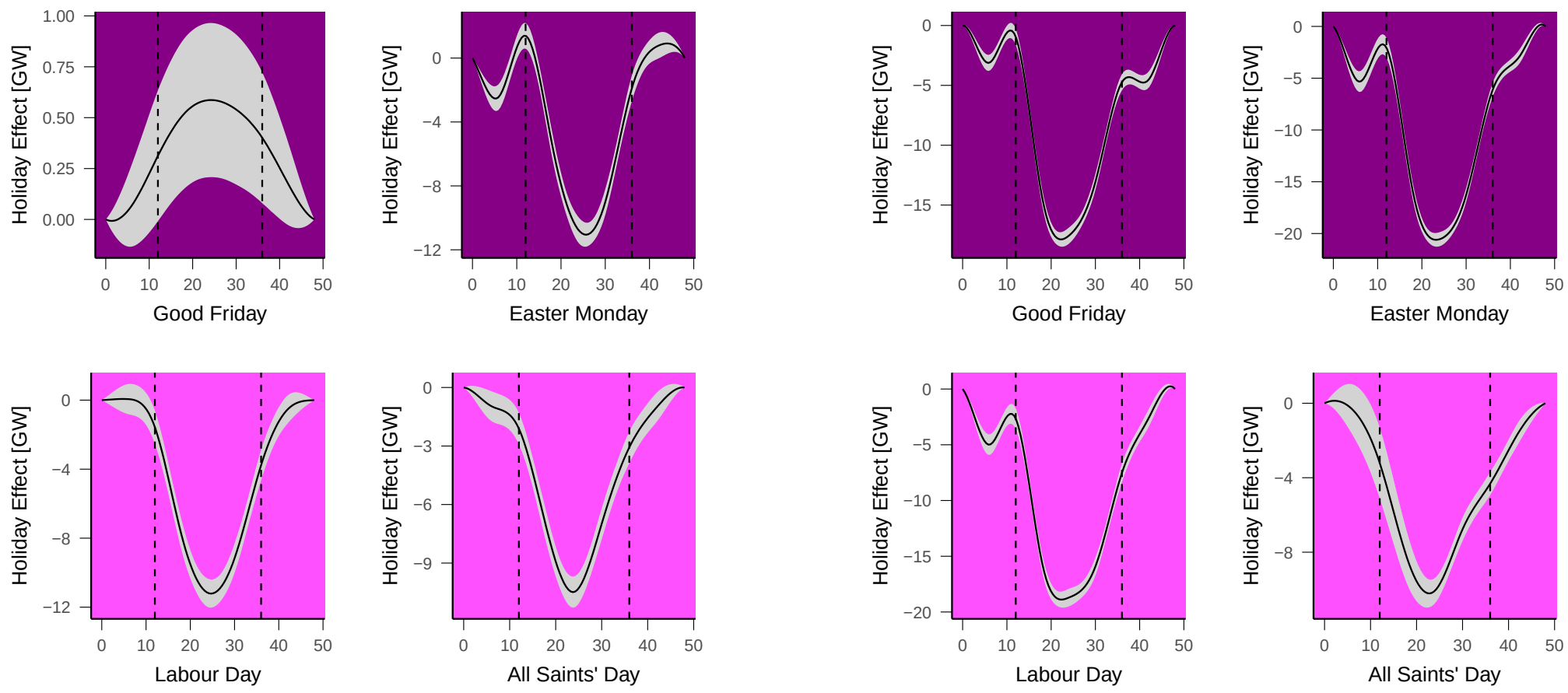

(a) France (b) Germany

Figure 12: The smooth effects of bridged fixed day (dark purple) and weekday (magenta) holidays by (7) on load with dashed lines indicating the first and last hour of the holiday.

In France (Figure 13), as noted before, a clear temperature effect is observed with varying degrees from year to year (maximal load increase approx. 22 GW in 2016-2017, 12 GW in 2017-2018), whereas the effect of the states is minimal (maximal approx. 2 GW). The days where the effect of the Christmas holiday period (maximal load decrease approx. 5 GW) are active and non-active are distinctly visible. For the fixed date holidays, we observe varying impacts, regarding the weekday of the holiday in the specific year (e.g. Christmas Day maximal load decrease approx. 13 GW (Fri) in 2015-2016, 0 GW (Sat) in 2016-2017, 13 GW (Mon) in 2017-2018). The effect of bridging the active hours of a holiday to the next day is particularly notable for the day after New Year's Day (e.g. maximal load decrease approx. 4 GW in 2017-2018).

In Germany (Figure 14), consistent with previous observations, the impact of temperature is minimal (maximal approx. 2 GW). Similarly, the mid-term states exhibit a negligible effect within the depicted timeframes (maximal approx. 2 GW). Accordingly, on non-holidays seasonal effects, including their interactions, are the primary driver of the load profile with a typical daily and weekly pattern. For the fixed date holiday Labour Day no effect is observed in 2016, where the holiday occurs on a Sunday. In 2015 and 2017 this holiday has a load-reducing effect of approx. 17 GW. Note that the minimal holiday effect observed on May 4th, 2016, is attributable to the bridged effect of Ascension Day on May 5th of the same year.

Similar plots for the Easter holiday time in France and the Christmas holiday time in Germany can be found in the Appendix B, see Figures 19, 18.

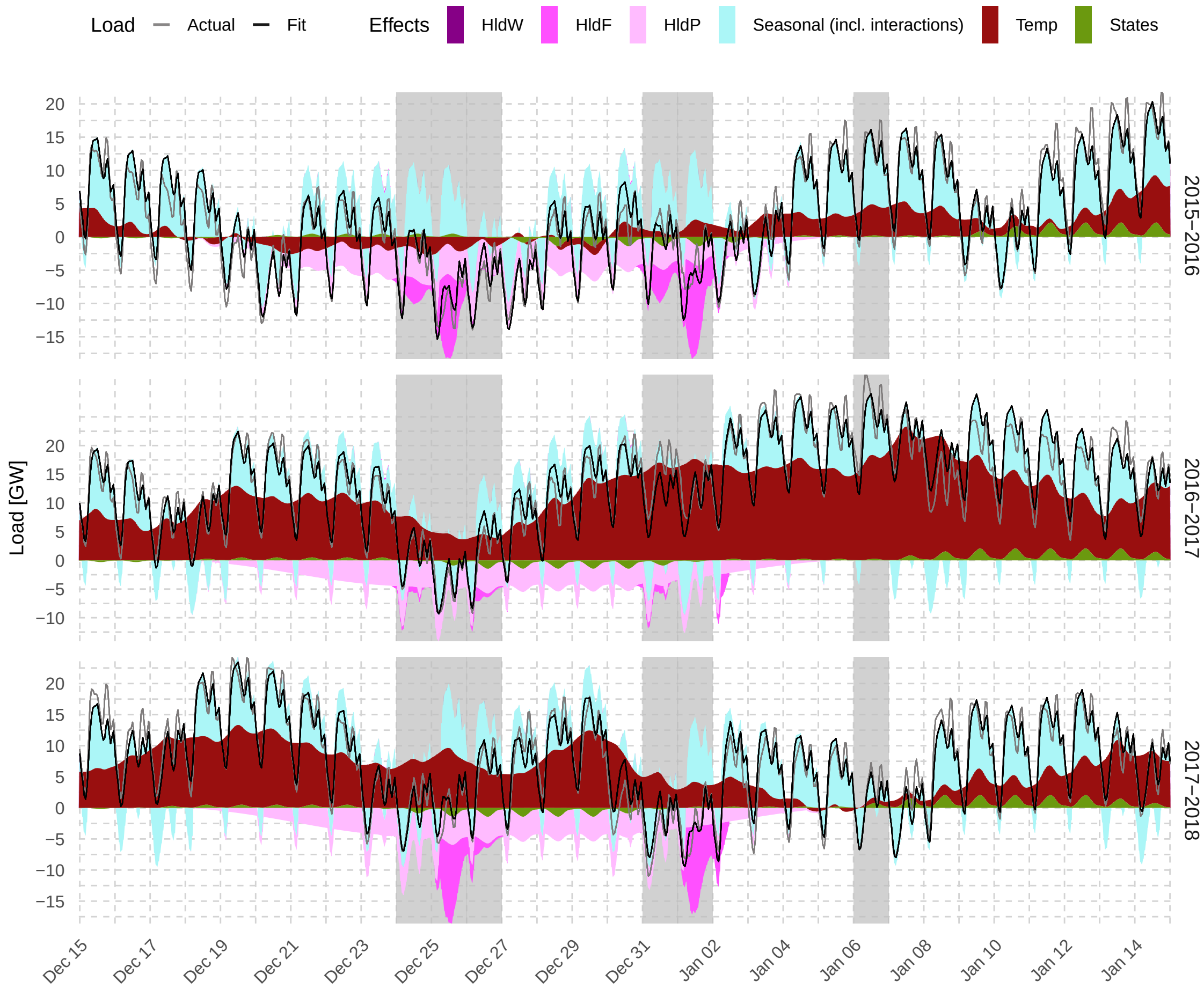

Figure 13: Estimated smooth effects, see (7), on load, fitted and actual load reduced by the estimated intercept in France from December 15th 2015, 2016 and 2017 to January 15th 2016, 2017 and 2018, with holidays shaded in grey.

## 7. Conclusion

We present a novel, two-stage GAM for effective, interpretable and robust mid-term hourly load forecasting across Europe. The model explicitly captures the multifaceted nature of load data by incorporating nuanced holiday modeling with vacation periods, fixed date and weekday holiday effects and exponentially smoothed temperatures. These inclusions significantly improve forecasting accuracy, as demonstrated in an evaluation across 24 European countries over 9 years (2015-2024) of data. Accounting for the socio-economic characteristics in the load time series through a persistent trend within the ETS framework proves challenging during the COVID-19 pandemic. Incorporating exogenous COVID-19 lockdown data improves model performance by capturing the sharp but brief load reduction during this period. Built as a linear model in interpretable P-splines, the model allows for understanding the impact of each incorporated component on load. Additionally, its linear structure enables remarkably fast computation times. These attributes, along with its significant improvements over standard linear models, make it a compelling candidate for model ensembling with neural networks.

Extensions of this work will apply the proposed model for probabilistic forecasting. Generating load simulations from our model is a straightforward process, as the model, constructed by independently estimated stages, encompasses independent error distributions. By this, meteorological uncertainties, which are crucial in risk assessment for mid-term power plant operations, energy storage evaluation and energy trading, can be incorporated.

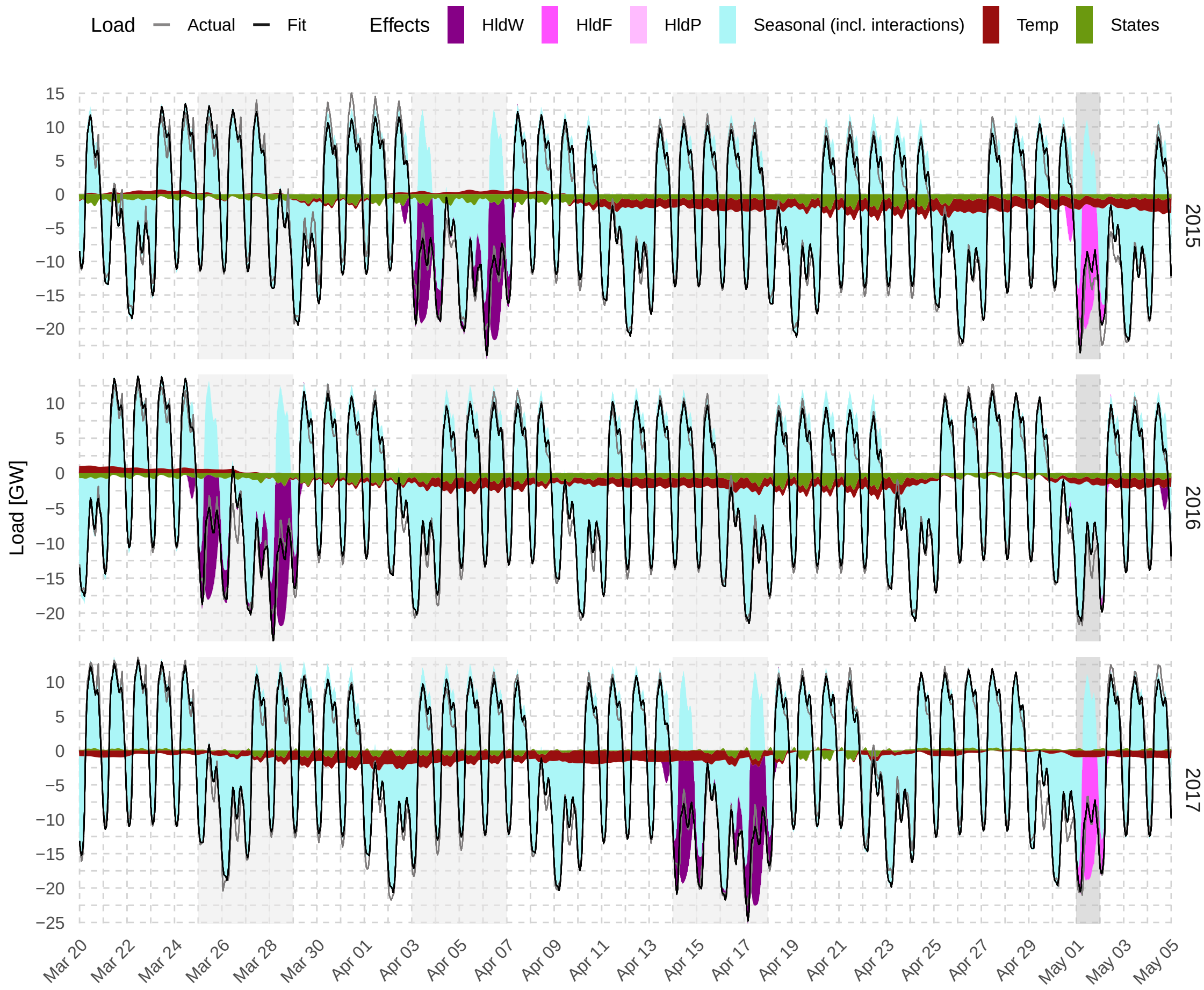

Figure 14: Estimated smooth effects, see (7), on load, fitted and actual load reduced by the estimated intercept in Germany from March 1st 2016, 2017 and 2018 to May 10th 2016, 2017 and 2018, with weekday holidays shaded in light grey and fixed date holidays shaded in dark grey.

### Author Contributions

M. Z.: Conceptualization; methodology; software; validation; formal analysis; investigation; data curation; writing - original draft; writing review and editing; visualization.

F. Z.: Conceptualization; methodology; validation; formal analysis; funding acquisition; resources; writing - review and editing; supervision.

### Funding Statement

This research has been funded by the Deutsche Forschungsgemeinschaft (DFG, German Research Foundation) – 505565850.

# Appendices

## A. Appendix: Tables

Table 8: Color-coded (red: low, blue: high) forecasting accuracy in terms of $\overline{\text{RMSE}(\widehat{\mathcal{E}}_h)}$ for $h = 1, ..., 168 \times 52$ in GW with the best model per country bolded. TSO day-ahead forecasts are included for countries with less than 1% of forecasts missing.

| | SRWW | SRWA | STL | VanDet | VanBas | VanRec | ARWD | FNN | CLIM | CL-ETS | GAM | \AR | \HLD | \Temp | \States | DA TSO |
|---|---|---|---|---|---|---|---|---|---|---|---|---|---|---|---|---|
| Austria | 1.091 | 0.627 | 0.646 | 1.291 | 1.189 | 1.181 | 0.768 | 0.494 | 0.510 | 0.508 | 0.501 | 0.501 | 0.561 | 0.523 | **0.450** | 0.444 |
| Belgium | 1.221 | 0.731 | 0.740 | 1.281 | 1.198 | 1.183 | 0.893 | 0.648 | 0.619 | 0.615 | 0.611 | 0.612 | 0.647 | 0.632 | **0.571** | – |
| Bulgaria | 0.982 | 0.458 | 0.502 | 0.805 | 0.649 | 0.633 | 0.687 | 0.381 | 1.587 | 0.403 | 0.380 | 0.380 | 0.393 | 0.397 | **0.371** | 0.188 |
| Czech Rep. | 1.286 | 0.583 | 0.613 | 1.272 | 1.113 | 1.096 | 0.887 | 0.452 | 0.473 | 0.463 | 0.451 | 0.452 | 0.530 | 0.492 | **0.407** | 0.178 |
| Denmark | 0.569 | 0.336 | 0.321 | 0.704 | 0.659 | 0.643 | 0.433 | 0.311 | 0.273 | 0.272 | **0.268** | 0.269 | 0.282 | 0.274 | 0.283 | 0.075 |
| Estonia | 0.198 | 0.091 | 0.095 | 0.186 | 0.150 | 0.149 | 0.145 | 0.072 | 0.075 | 0.075 | 0.074 | 0.074 | 0.077 | 0.078 | **0.070** | 0.091 |
| Finland | 1.962 | 1.003 | 0.899 | 1.454 | 0.929 | 0.933 | 1.459 | 0.758 | 0.742 | 0.759 | 0.731 | 0.732 | 0.745 | 0.773 | **0.714** | 0.225 |
| France | 14.771 | 6.305 | 6.257 | 10.181 | 7.145 | 6.977 | 10.344 | 5.094 | 5.407 | 5.277 | 5.037 | 5.061 | 5.103 | 5.825 | **4.779** | 1.243 |
| Germany | 7.157 | 4.544 | 4.369 | 9.519 | 9.203 | 9.047 | 5.171 | 3.583 | 3.095 | 3.027 | **3.007** | 3.024 | 3.682 | 3.136 | 3.071 | 2.645 |
| Greece | 1.251 | 0.754 | 0.801 | 1.197 | 1.079 | 1.041 | 0.870 | 0.610 | 0.628 | 0.622 | 0.603 | 0.609 | 0.623 | 0.659 | **0.595** | 0.246 |
| Hungary | 0.719 | 0.439 | 0.472 | 0.741 | 0.696 | 0.683 | 0.514 | 0.350 | 0.369 | 0.382 | 0.377 | 0.378 | 0.411 | 0.381 | **0.323** | 0.269 |
| Italy | 5.384 | 3.524 | 4.140 | 7.419 | 7.288 | 7.155 | 3.753 | 2.912 | 2.988 | 2.953 | 2.930 | 2.943 | 3.359 | 3.115 | **2.524** | 0.901 |
| Latvia | 0.109 | 0.067 | 0.064 | 0.154 | 0.144 | 0.143 | 0.085 | 0.060 | 0.047 | 0.048 | 0.047 | **0.046** | 0.053 | 0.049 | 0.052 | 0.026 |
| Lithuania | 0.198 | 0.116 | 0.123 | 0.251 | 0.233 | 0.231 | 0.144 | 0.097 | 0.100 | 0.101 | 0.100 | 0.099 | 0.108 | 0.102 | **0.090** | 0.064 |
| Montenegro | 0.087 | 0.056 | 0.053 | 0.084 | 0.078 | 0.073 | 0.066 | 0.050 | 0.050 | 0.048 | 0.047 | 0.047 | **0.047** | 0.053 | 0.047 | 0.037 |
| Netherlands | 2.015 | 1.253 | 1.627 | 1.964 | 1.868 | 1.844 | 1.604 | 1.371 | 1.175 | 1.191 | 1.188 | 1.182 | 1.208 | **1.159** | 1.276 | 1.775 |
| Poland | 2.454 | 1.550 | 1.567 | 3.208 | 3.097 | 3.050 | 1.693 | 1.063 | 1.092 | 1.055 | 1.032 | 1.035 | 1.357 | 1.089 | **0.935** | 0.677 |
| Portugal | 0.769 | 0.452 | 0.553 | 1.006 | 1.012 | 0.975 | 0.534 | 0.384 | 0.411 | 0.409 | 0.405 | 0.409 | 0.440 | 0.417 | **0.329** | 0.226 |
| Romania | 0.930 | 0.615 | 0.622 | 1.028 | 0.944 | 0.944 | 0.691 | 0.505 | 0.493 | 0.487 | 0.480 | 0.483 | 0.523 | 0.509 | **0.457** | 0.129 |
| Serbia | 0.873 | 0.467 | 0.477 | 0.777 | 0.624 | 0.630 | 0.695 | 0.504 | 0.421 | 0.419 | **0.409** | 0.412 | 0.416 | 0.437 | 0.452 | 0.200 |
| Slovakia | 0.459 | 0.321 | 0.273 | 0.496 | 0.457 | 0.458 | 0.348 | 0.263 | 0.239 | 0.233 | **0.230** | 0.231 | 0.252 | 0.244 | 0.239 | 0.118 |
| Slovenia | 0.259 | 0.159 | 0.156 | 0.294 | 0.280 | 0.278 | 0.191 | 0.141 | 0.127 | 0.126 | **0.125** | 0.126 | 0.141 | 0.130 | 0.131 | 0.068 |
| Spain | 3.601 | 2.326 | 2.711 | 4.404 | 4.323 | 4.224 | 2.643 | 2.094 | 2.107 | 2.091 | 2.046 | 2.065 | 2.241 | 2.101 | **1.925** | 0.427 |
| Sweden | 4.102 | 1.586 | 1.590 | 3.073 | 1.910 | 1.906 | 2.941 | 1.233 | 1.264 | 1.277 | 1.226 | 1.229 | 1.252 | 1.286 | **1.200** | 0.532 |
| Sum | 52.448 | 28.362 | 29.673 | 52.788 | 46.268 | 45.475 | 37.557 | 23.430 | 24.290 | 22.841 | 22.306 | 22.397 | 24.452 | 23.861 | **21.290** | 10.787 |

## B. Appendix: Figures

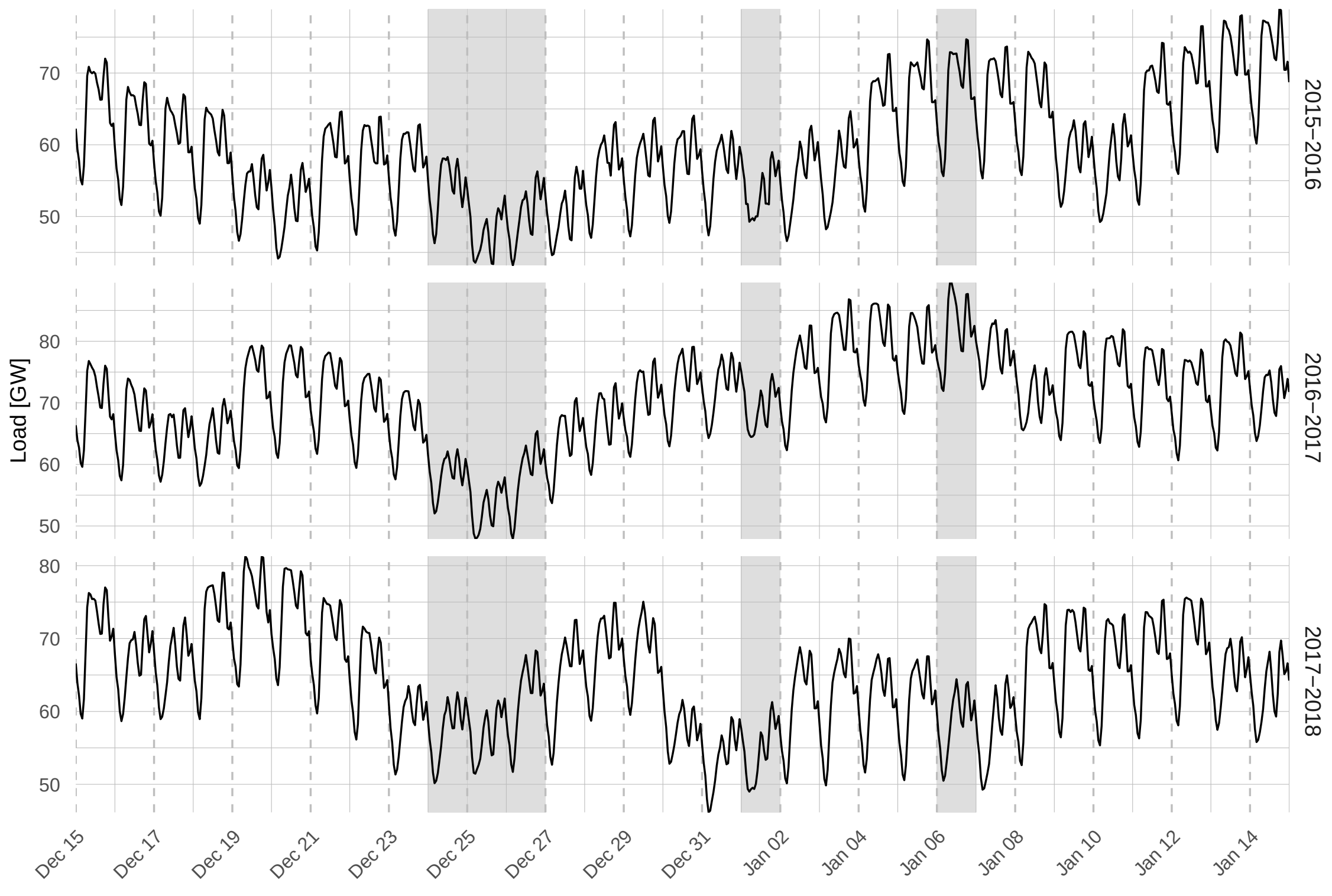

(a) December 15th in 2015, 2016 and 2017 to January 15th in 2016, 2017 and 2018.

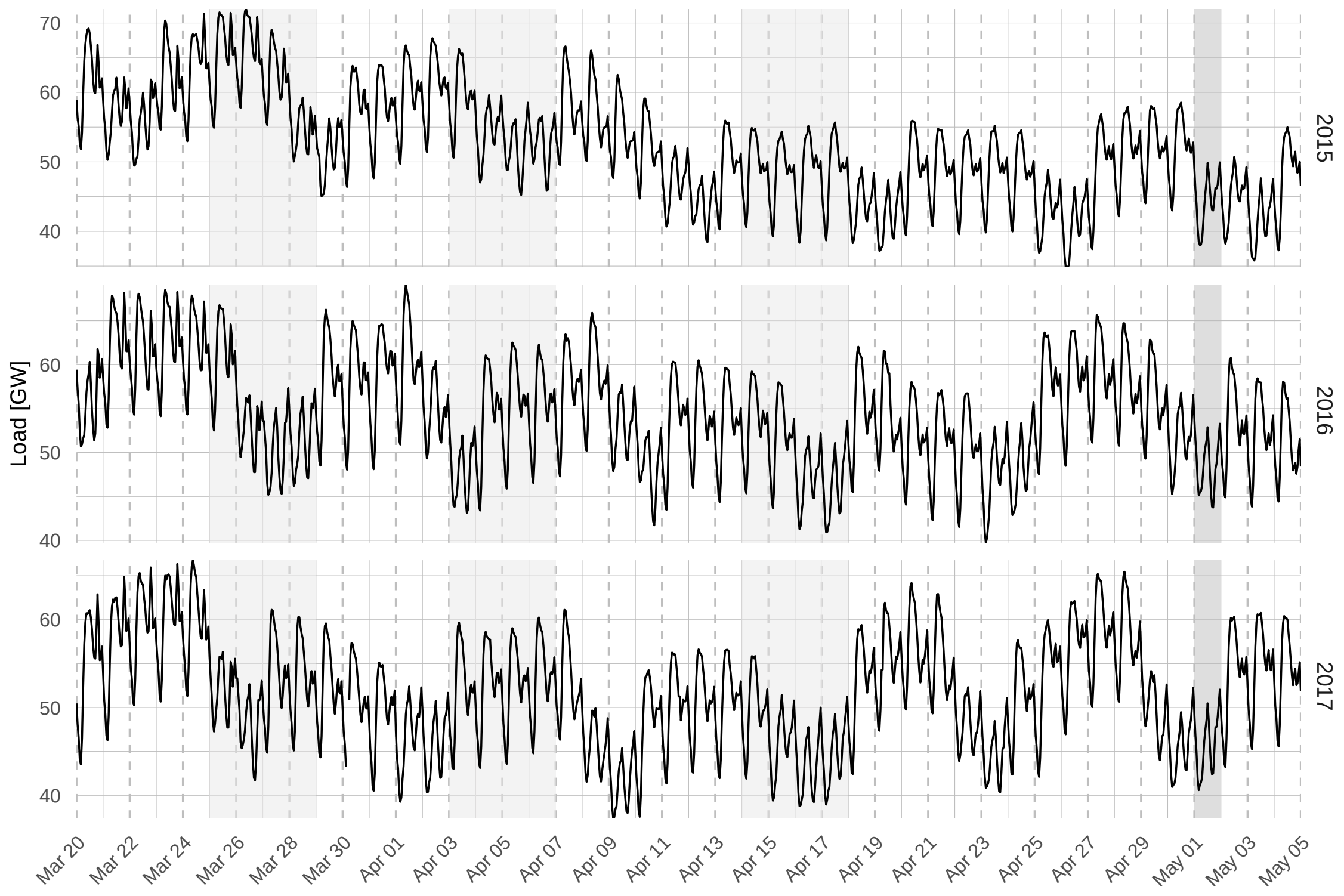

(b) March 1st to May 10th in 2016, 2017 and 2018.

Figure 15: Hourly load in France in the Easter and Christmas holiday time with weekday holidays shaded in light grey and fixed date holidays shaded in dark grey.

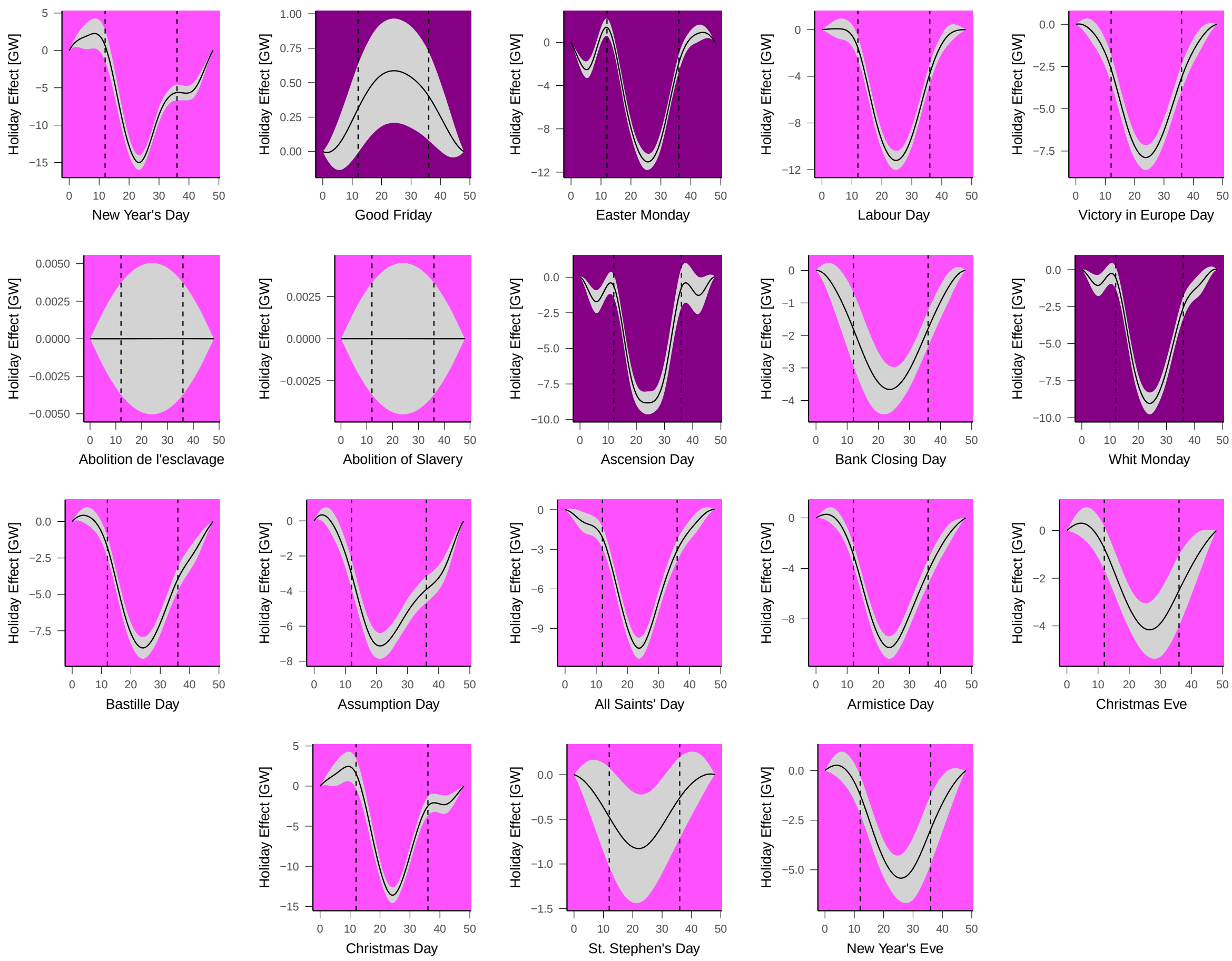

Figure 16: The smooth effects of bridged fixed day (purple) and weekday (light purple) holidays by (7) on load in France with dashed lines indicating the first and last hour of the holiday..

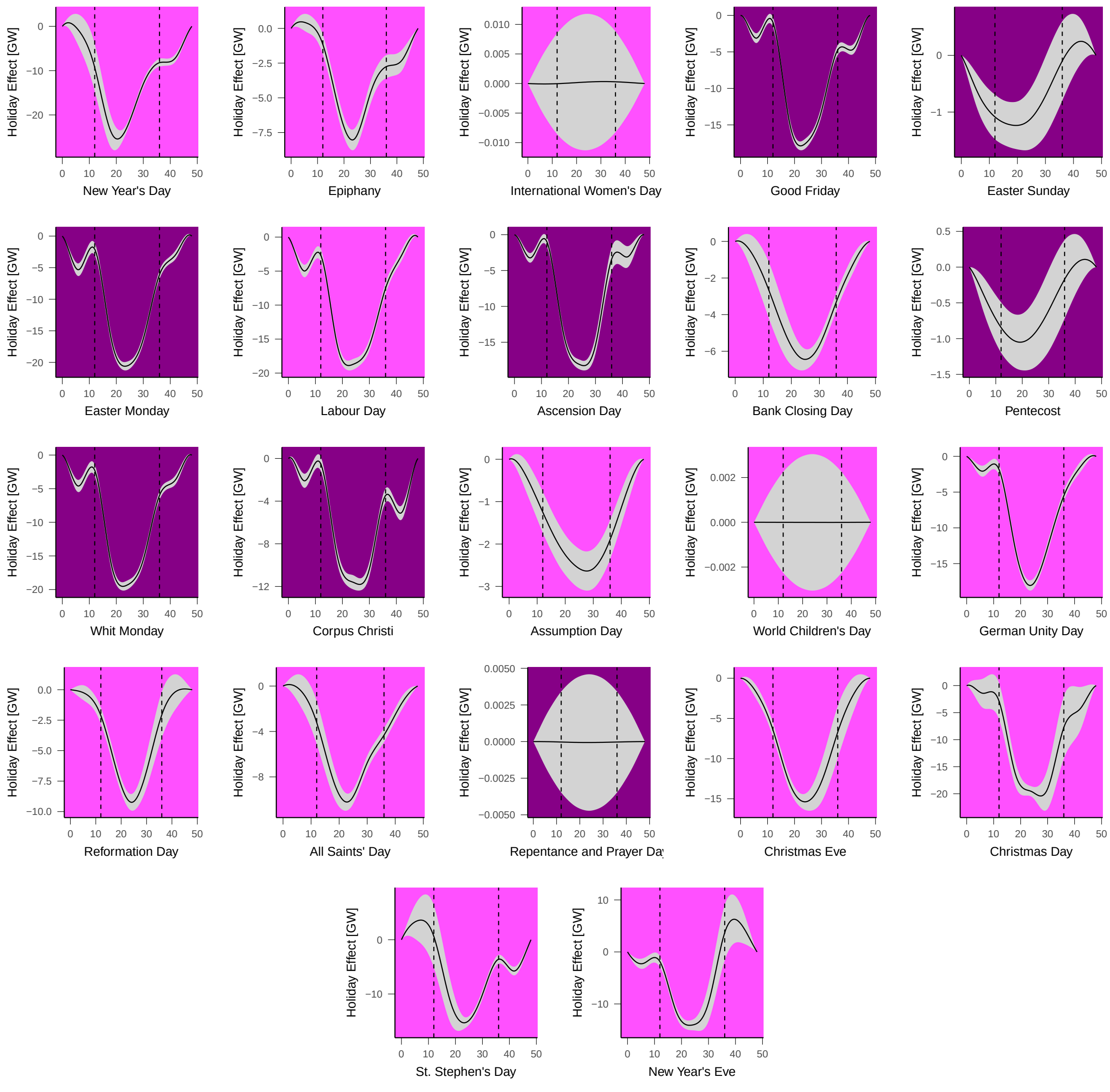

Figure 17: The smooth effects of bridged fixed day (purple) and weekday (light purple) holidays by (7) on load in Germany with dashed lines indicating the first and last hour of the holiday.

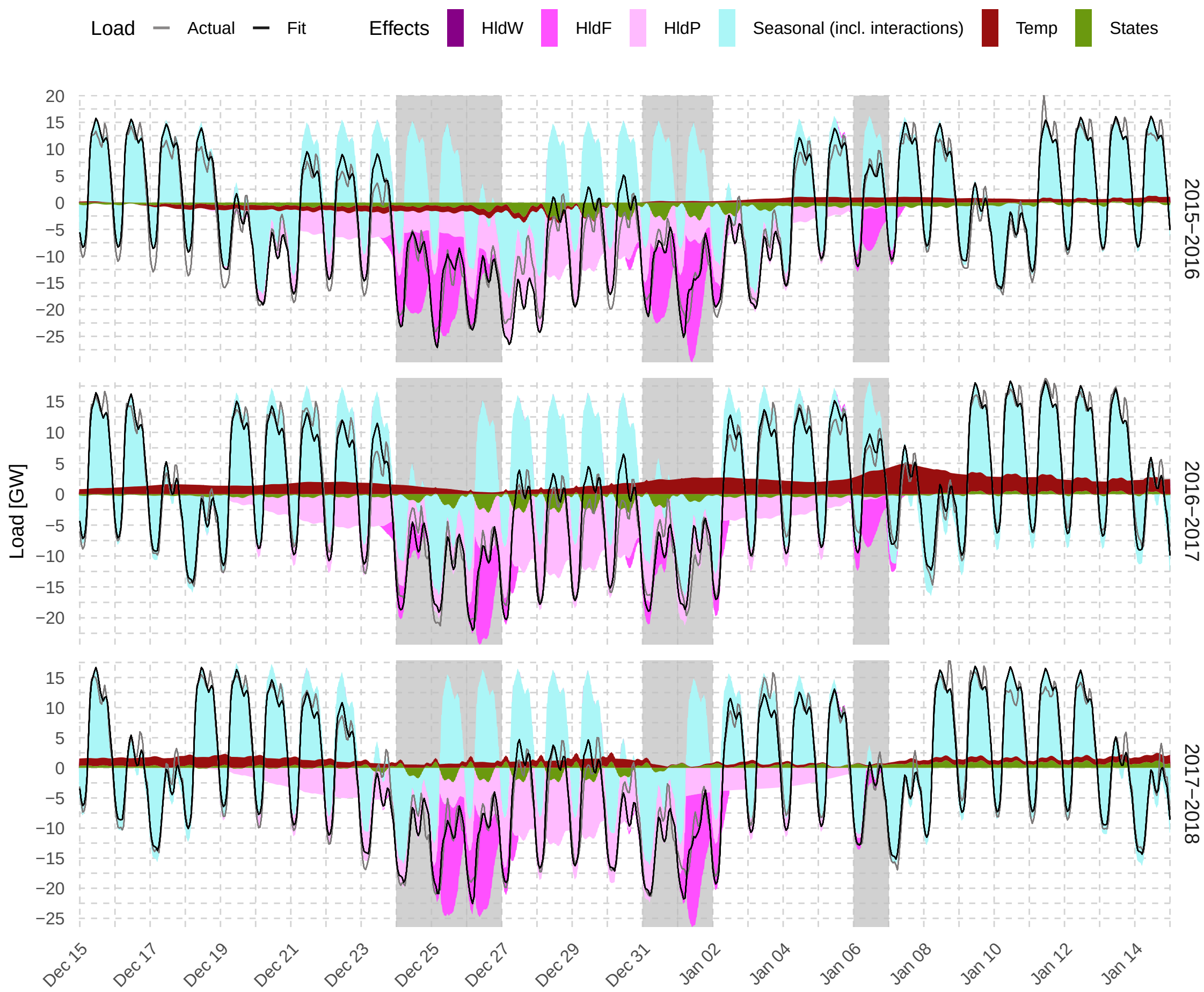

Figure 18: Estimated smooth effects, see (7), on load, fitted and actual load reduced by the estimated intercept in Germany from December 15th 2015, 2016 and 2017 to January 15th 2016, 2017 and 2018, with holidays shaded in grey.

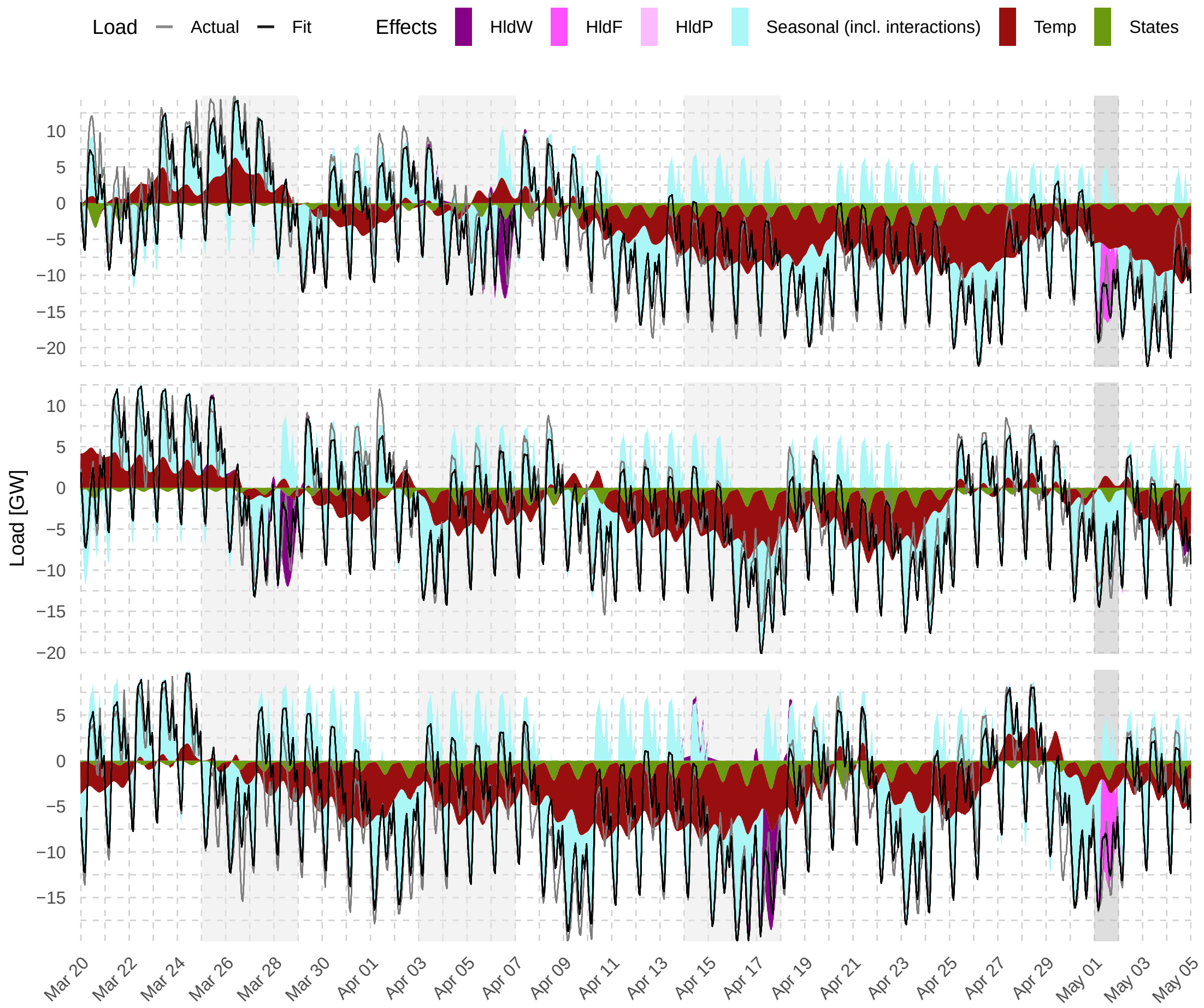

Figure 19: Estimated smooth effects, see (7), on load, fitted and actual load reduced by the estimated intercept in France from March 1st 2016, 2017 and 2018 to May 10th 2016, 2017 and 2018, with weekday holidays shaded in light grey and fixed date holidays shaded in dark grey.